\documentclass[12pt]{iopart}

\usepackage{graphicx}

\newcommand\nc[2]{\newcommand#1{#2}}
\nc{\ee[1]}{\hbox{\sl e}^{\, #1}}
\nc{\ordo[1]}{ {\cal O} \hskip -.1ex \left( #1 \right) }
\nc\tbeg{t_{\hbox{\scriptsize beg}}}
\nc\tend{t_{\hbox{\scriptsize end}}}
\nc\qd{{\rm d}}
\nc{\new[1]}{{\sl #1}}

\begin{document}

\title{Boundary effect of a partition in a quantum well}

\author{T F\"ul\"op$^1$%
\footnote{Postdoctoral fellow at the Doppler Institute for Mathematical
Physics and Applied Mathematics, Prague, Czech Republic}
and I Tsutsui$^2$}

\address{$^1$ Nuclear Physics Institute, Academy of Sciences, 25068
\v Re\v z near Prague, Czech Republic}
\ead{tamas.fulop@gmail.com} 

\address{$^2$ Institute of Particle and Nuclear Studies, High Energy
Accelerator Research Organization (KEK), Tsukuba 305-0801, Japan} 
\ead{izumi.tsutsui@kek.jp} 

\begin{abstract}
The paper wishes to demonstrate that, in quantum systems with boundaries,
different boundary conditions can lead to remarkably different physical
behaviour. Our seemingly innocent setting is a one dimensional potential
well that is divided into two halves by a thin separating wall. The two
half wells are populated by the same type and number of particles and are
kept at the same temperature. The only difference is in the boundary
condition imposed at the two sides of the separating wall, which is the
Dirichlet condition from the left and the Neumann condition from the
right. The resulting different energy spectra cause a difference in the
quantum statistically emerging pressure on the two sides. The net force
acting on the separating wall proves to be nonzero at any temperature
and, after a weak decrease in the low temperature domain, to increase and
diverge with a square-root-of-temperature asymptotics for high
temperatures. These observations hold for both bosonic and fermionic type
particles, but with quantitative differences. We work out several
analytic approximations to explain these differences and the various
aspects of the found unexpectedly complex picture.
\end{abstract}
\pacs{03.65.-w, 02.30.Mv, 02.30.Tb, 02.60.-x, 02.60.Lj, 05.30.-d,
05.30.Jp, 05.30.Fk}
\submitto{\JPA}
\maketitle

\section{Introduction}
\label{s1}

Quantum singularity is a point defect in an otherwise `regular' system
treated in quantum mechanics. Despite the simple setting, such a system
exhibits various features which are intriguing both mathematically and
physically. On mathematical sides, it permits rich structures admitted
under the general class of self-adjoint Hamiltonians describing the
system. The mathematically allowed class is rather large, but when the
system is linear ({\it i.e.}, one dimensional), it is concisely
characterized by the unitary group $U(2)$ (see, {\it e.g.},
\cite{RS,AG,AGHH}) which includes, among others, the familiar singular
interactions realized by the Dirac delta potentials with arbitrary
strengths. On physical sides, one the other hand, the quantum singularity
admits interesting phenomena such as duality, supersymmetry, anholonomy
(Berry phase) and spontaneous symmetry breaking \cite{CFT,Moebius,FT}
(see also \cite{Exner}). An application for a qubit device, which might
become realizable by future development of nanotechnology, has also been
suggested \cite{CTF}.

All the studies on quantum singularity mentioned above have assumed
one-particle systems at zero temperature, but since actual realizations
of the systems would involve many particles acting under finite
temperatures, it should be inevitable to take into account the
statistical aspects of the particles. The aim of this paper is to present
a case study of such statistical aspects by analyzing a system which is
both simple and familiar -- a quantum well. Specifically, we consider
particles in a quantum well with a partition placed in the centre. The
partition is made by an infinitely thin wall which forms a single point
defect inside the well. The partition separates the well into two half
wells, each of which is supposed to contain $N$ identical particles. For
the statistics of the particles, we consider separately the two cases,
the Bose-Einstein and the Fermi-Dirac statistics, and the entire well is
put under various temperatures, ranging from zero to infinity.

\begin{figure}
\centering
\resizebox{.4\textwidth}{!}{\includegraphics{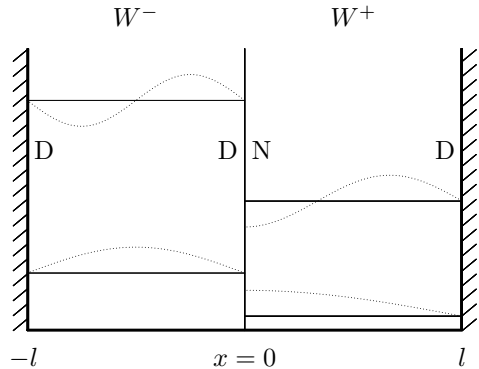}}
\caption{
Quantum well with a partition at the centre. Under the Dirichlet ($ \psi
= 0 $) and Neumann ($ \psi' = 0 $) boundary conditions on the left and
the right of the partition (while we choose the Dirichlet for the two
ends of the well), the eigenstates (the lowest two are shown in both half
wells $W^\pm$) have different energy levels. When the same number $N$ of
particles are introduced in each of the half wells, these level
differences lead to a net force on the partition which exhibits
interesting temperature behaviours.
}\label{fig:1}
\vskip 3.4ex
\end{figure}

To characterize the partition in a simplest nontrivial term, we assume
that the partition enforces the Dirichlet condition on the left and the
Neumann on the right (see figure~\ref{fig:1}), which is perhaps a simplest
nontrivial combination permitted quantum mechanically. We focus on the
pressure, or statistical quantum force, acting on the partition, which
arises from the discordance in energy levels between the two half wells
\cite{Berman}. We shall be content to work in the one dimensional setup,
but the extension to multi-dimensions will be done analogously by
considering the class of singularities pertinent to the dimensions.

The temperature behaviour of the net quantum force --- the difference
between the pressures from the left and the right of the partition ---
exhibits a number of interesting and unexpected features. For instance,
the force curve as a function of temperature starts off with a finite
value at the zero temperature and reaches a single minimal point at a
finite temperature before diverging in the high temperature limit
according to a certain scaling law, {\it i.e.}, the square root of the
temperature. This overall temperature behaviour can be observed for both
the bosonic and the fermionic cases, but in the latter case the curve in
the low temperature regime shows an additional little twist which is
absent in the former. Another feature worth mentioning is the distinct
characteristics of the dependence on the number of particles; {\it e.g.},
both the zero temperature value and the minimal force are of the order of
$N$ for bosons while they are of the order of $N^2$ for fermions.

Our analysis on the quantum force is carried out both numerically and
analytically. Some of the initial results have been reported earlier
\cite{FMT,TF}, and here we shall present the full detail of our complete
analysis, including the analytical approximations performed with
assistance of numerical solutions of some transcendental equation. We
shall see that our goal to find analytic formulae that can account for
the pure numerical results as well as the statistical feature of the
particle number dependence mentioned above is achieved reasonably well.
Our outcomes illustrate how quantum singularity, realized by a partition
in the present model of a quantum well, can give rise to physically
measurable effects in a plain form such as pressure.

This paper is organized as follows.  In section~\ref{s2} we provide the
basic account of our model as well as the overall features of the quantum
force obtained by numerical computations for all temperature regions. In
section~\ref{s3}, we present the analysis for the high temperature limit
where the force exhibits the common scaling behaviour for the two types of
particle statistics. Section~\ref{s4} is devoted to the analysis of the
medium temperature regime where the force curve has a minimal point. Our
argument for the analytic approximation is given for bosons and fermions,
separately, since the statistical difference becomes important in this
regime. Similarly, section~\ref{s5} is devoted to the analysis of the low
temperature regime where the force possesses the characteristics of the
$N$-dependence and further develops the twist in the fermionic case. In
section~\ref{s6} we briefly mention two physical quantities, the shift of
partition and the transfer of particles between the two half wells, as
alternatives which may be more directly observed than the quantum force.
Finally, we present our conclusion and discussions in section~\ref{s7}.
Appendix~\ref{appa} gives the outline of our numerical computation, and
Appendix~\ref{appb} contains the technical detail of our approximation
employed in section~\ref{s4}.

\section{Quantum force on a partition and its temperature behaviours}
\label{s2}

Before we present our full analysis of the quantum force, we provide the
basic account of the system for which we discuss the boundary effect for
various temperature regions.

The system we consider is a one-dimensional quantum well $W$ possessing a
partition wall at the centre, given by the interval $[-l, l]$ with the
partition at $x = 0$. To define the system in quantum mechanics, we need
to specify the boundary conditions for the wave function $\psi(x)$
imposed both at the two ends of the well $x = \pm l$ and at the partition
on its left and right sides $x = \pm 0$. In quantum mechanics, allowed
boundary conditions are those which respect the probability conservation
requirement. If we assume that the partition is impenetrable, {\it i.e.},
the particles cannot penetrate the partition, then the most general form
of the boundary conditions is given by
  \begin{equation}
  \psi(\pm l) + L_{\pm l}\,  \psi'(\pm l) = 0,  \qquad  
  \psi(\pm 0) + L_{\pm 0}\,  \psi'(\pm 0) = 0, 
  \label{eq:bcgen}
  \end{equation}  
where $L_{\pm l}$, $L_{\pm 0}$ are arbitrary real constants including
infinity. For instance, $L_{+0} = 0$ in (\ref{eq:bcgen}) implies the
Dirichlet condition $\psi(+0) = 0$ whereas $L_{+0} = \infty$ implies the
Neumann condition $\psi'(+0) = 0$ at $x = +0$. Similar choices of
boundary conditions should be made at the other three points, $x = \pm l$
and $x = -0$ by specifying the constants $L_{\pm l}$ and $L_{-
0}$.\footnote{
The allowed set of boundary conditions (\ref{eq:bcgen}) forms the group
$[U(1)]^4$ representing the possible self-adjoint domains of the free
Hamiltonian in the interval $[-l, l]$ divided into two. If we allow the
partition to transmit particles, then the boundary condition can be more
general than (\ref{eq:bcgen}) and given by $U(2) \times [U(1)]^2$ (see
\cite{RS,AG,AGHH}).
}

\begin{figure}
\centering
\resizebox{.35\textwidth}{!}{\includegraphics{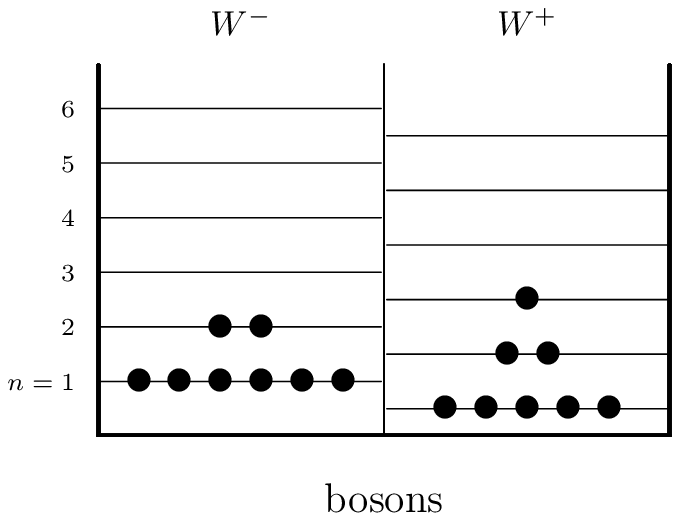}}
\hskip .2\textwidth
\resizebox{.35\textwidth}{!}{\includegraphics{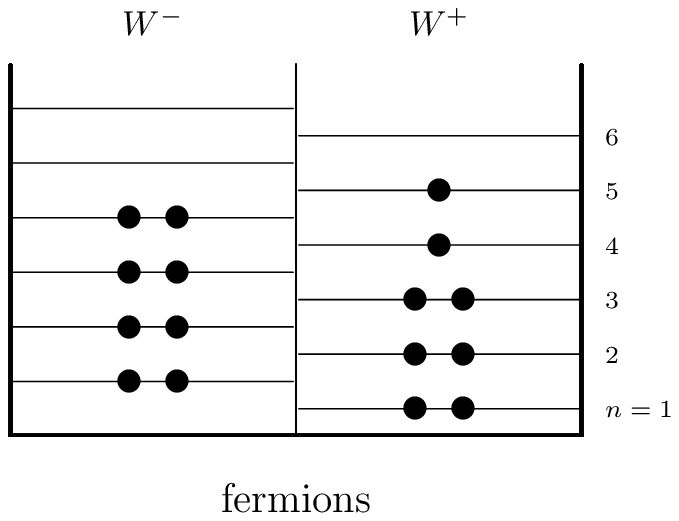}}
\caption{
Illustration of the distribution of particles over the levels in the two
half wells at a low temperature. In the fermionic example $s = 1/2$ is
assumed. The gap of the levels and the difference in the populations in
the two wells induce a non-vanishing force on the partition, which is
dependent on the temperature as well as the particle statistics.
}\label{fig:2}
\vskip 3.4ex
\end{figure}

In this paper, to realize a nontrivial partition in the simplest setting,
we adopt the set of boundary conditions provided by $L_{\pm l} = 0$,
$L_{+0} = \infty$ and $L_{-0} = 0$, that is, the Neumann boundary
condition at the right side $x = +0$ and the Dirichlet at the left side
$x = -0$ of the partition as well as at the both ends $x = \pm l$ of the
wall:
  \begin{equation}
  \psi (-l) = 0, \qquad \psi(-0) = 0, \qquad
  \psi'(+0) = 0, \qquad \psi (l) = 0.
  \label{eq:bcond}
  \end{equation}
The quantum well $W$ is then split into two half, physically distinct,
wells $W^+$ and $W^-$ by the partition. In the two half wells $W^\pm$ with
the boundary conditions (\ref{eq:bcond}), the free Hamiltonian
  \begin{equation}
  H = -\frac{\hbar^2}{2m}\frac{\qd^2}{\qd x^2}
  \label{eq:noH}
  \end{equation} 
admits energy levels 
  \begin{equation}
  E_n = e_n {\cal E}, \qquad 
  {\cal E} = \frac{\hbar^2}{2m}\left(\frac{\pi}{l}\right)^2,
  \qquad  n = 1, 2, 3 \ldots,
  \label{eq:Qunitenergy}
  \end{equation}
which are distinct for the two half wells (see figure~\ref{fig:1}),
  \begin{equation}
  e_n = \left\{ \begin{array}{cl}  \left( n - \frac{1}{2} \right)^2 &
  \mbox{ for } W^+ ,  \\  n^2 & \mbox{ for } W^- .  \end{array} \right.
  \label{eq:unitenergy}
  \end{equation}
It is then expected that the level gap between the two half wells $W^\pm$
gives rise to noticeable physical effects, and one obvious example will
be the force (or pressure) acting on the partition due to the
difference of the pressures exerted by particles inside the two half
wells. To extract the pure boundary effect, we put an identical number
of particles in each of the half wells, and thereby study the temperature
dependence of the net force emerging on the partition. We do this for the
two cases of particle statistics, bosons and fermions, which possess
different features because of the different statistical distributions
over the energy levels (see figure~\ref{fig:2}).

To be more explicit, let $N$ denote the number of the identical particles
in both of the two half wells $W^\pm$ in each of the $2s+1$ spin degrees
of freedom, where $s$ is the spin of the particles. At temperature $T$,
the particles distribute over the levels according to the population
number (per spin degree of freedom)
  \begin{equation}
  N_n = \frac{1}{e^{\alpha + b e_n} - \eta},
  \label{eq:popl}
  \end{equation}
where we have introduced the statistical index,
  \begin{equation}
  \eta = (-1)^{2s} = \left\{ \begin{array}{rl} 1 & \mbox{for bosons} \\
  -1 & \mbox{for fermions} \end{array} \right.
  \label{eq:fbindex}
  \end{equation}
and the shorthand 
  \begin{equation}
  b = \frac{{\cal E}}{k_B T}
  \label{eq:nob}
  \end{equation}
with $k_B$ being the Boltzmann constant and ${\cal E}$ given in
(\ref{eq:Qunitenergy}). The temperature-dependent quantity $\alpha$ is
determined by the total particle number constraint,
  \begin{equation}
  N = \sum_{n} N_n.
  \label{eq:constr}
  \end{equation}

The force acting on the partition from one side is then given by
  \begin{equation}
  F = - (2s + 1) \sum_n  N_n \frac{\partial E_n}{\partial l} =
  (2s + 1) \frac{2{\cal E}}{l} \sum_n N_n e_n,
  \label{eq:eqnQexpress}
  \end{equation}
which is simplified as
  \begin{equation}
  f = \sum_n N_n e_n,
  \label{eq:eqnQfsum}
  \end{equation}
in terms of the reduced (dimensionless) force defined by
  \begin{equation}
  f =\frac{1}{2s+1} \frac{l}{2{\cal E}} F.
  \label{eq:redf}
  \end{equation}
Denoting by $f^\pm$ the forces on the partition in the half wells
$W^\pm$, we have the net force on the partition (from the left to the
right),
  \begin{equation}
  \Delta f = f^{-} - f^{+} = \sum_n N_n^- e_n^- - \sum_n N_n^+ e_n^+.
  \label{eq:Qnfp}
  \end{equation}
Throughout the paper we use the superscripts $\pm$ to refer to quantities
pertaining to the half wells $W^\pm$, respectively, and also use $\Delta$
to refer to the difference between the two quantities such as
(\ref{eq:Qnfp}). However, we will often omit these superscripts for
brevity as long as the distinction is unnecessary in the argument.

\begin{figure}
\centering
\resizebox{.37\textwidth}{!}{\includegraphics{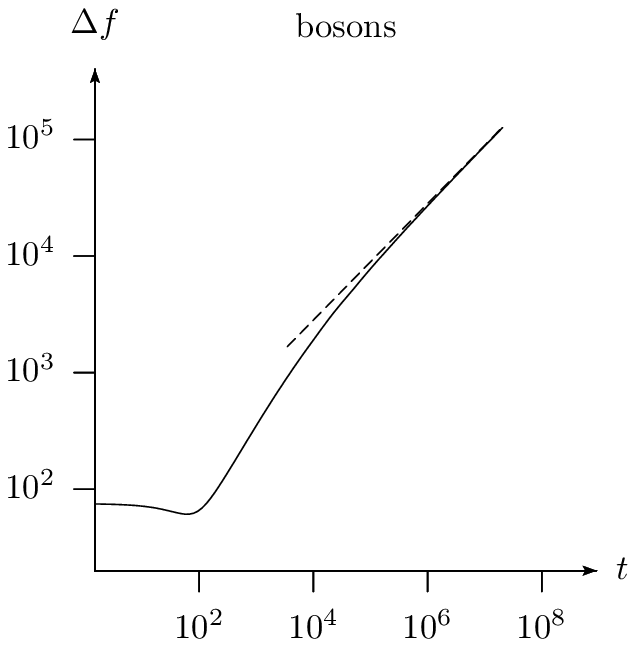}}
\hskip .1\textwidth
\resizebox{.394\textwidth}{!}{\includegraphics{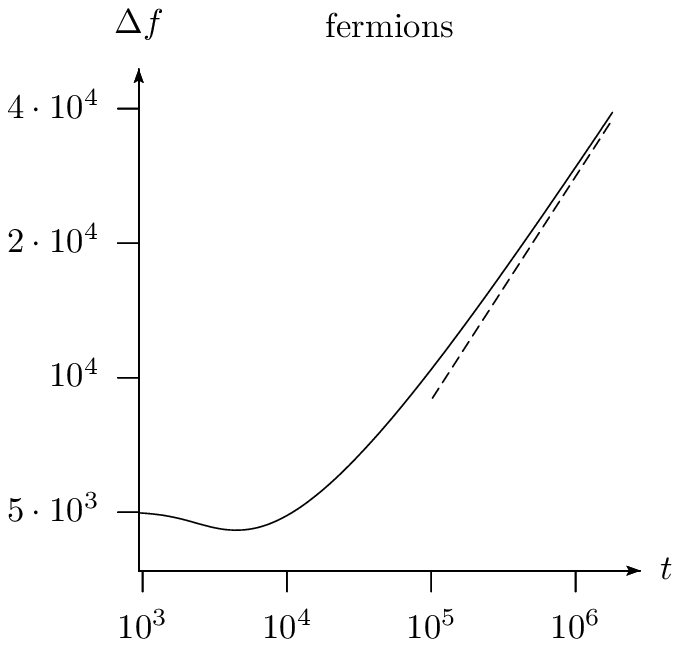}}
\caption{
Double logarithmic plot of the net force $\Delta f$ as the function of
the temperature variable $t$, for bosons (left) and fermions (right), at
$N = 100$, obtained by a numerical computation (solid line). Their common
high-temperature asymptotics (\ref{eq:eqnQaaj}) is also displayed (dashed
line).
}\label{fig:3}
\vskip 3.4ex
\end{figure}

Our aim will be to determine the net force $\Delta f$ in (\ref{eq:Qnfp})
as a function of the temperature. For our later convenience, we introduce
the reduced temperature $t$ by
  \begin{equation}
  t = \frac{1}{b} = \frac{k_B}{{\cal E}}T.
  \label{eq:redtemp}
  \end{equation}
Numerical evaluations of  $\Delta f$ in terms of $t$ can be achieved
readily for finite $N$, and the outcome for $N = 100$ is shown in
figure~\ref{fig:3} for the whole range of $t$.  We observe immediately
there that ({\rm i}) the net force has a finite zero temperature limit,
({\rm ii}) it exhibits a power law (as it approaches the dotted straight
line) in the high temperature limit, and ({\rm iii}) it has a single
minimum in between.
These overall features are common both in the bosonic and fermionic
cases, and can be seen for other values of $N$ different from $N = 100$ as
well. The basic difference between the bosonic and fermionic cases lies
in the fact that the strength of the force is of the order of $N$ for bosons while it
is of the order of $N^2$ for fermions.  Interestingly, the same difference in the $N$
dependence can also be seen for the temperature value of the minimum force. 
In most part of our
discussions, the numerical results will be presented for $N = 100$
particles in each spin degree of freedom (which is already a realistic
population number in nanoscale quantum experiments \cite{Nature}), but
the conclusion remains the same for larger $N$.

Analytical evaluations of the force, in contrast, are not readily
obtained, since the required sums in (\ref{eq:redf}) cannot be performed
exactly, and an analytical solution for $\alpha$ which has to be
determined is also difficult to find. We are thus forced to apply some
approximations, but the standard approach of approximating the sum with
an integral turns out to be insufficient to get the accuracy needed in
the present situation, leaving us to work out novel methods for the
present problem. For instance, the Fermi-Dirac integral, which is often
used to approximate the sum and is related to the Lerch transcendent, can
be expressed via an asymptotic series \cite{Stoner}, but the truncations
of the series do not provide enough precision to recover the force
difference obtained numerically. As we shall see soon, our approximations
developed in this paper and performed partly with the help of numerical
solutions for transcendental equations provide formulae which fit the
numerical results reasonably well. In particular, the three salient
features ({\rm i}), ({\rm ii}) and ({\rm iii}) mentioned above will be
seen to be all reproduced properly by our analytic methods.

\section{High temperature regime}
\label{s3}

As we have seen from the numerical results mentioned in figure~\ref{fig:3},
in the high-temperature limit the force $\Delta f$ exhibits a certain
scaling behaviour which is common to both the bosonic and fermionic cases.
We first show that this scaling behaviour can be explained analytically
based on a rather simple argument which is completely analogous in the
two cases.

To study the high-temperature regime, we first note that the population
number $N_1$ decreases for increasing temperature (see (\ref{eq:popl}) for
$n = 1$) and, accordingly, we expect $\alpha$ to increase to higher
positive values for large $t$ (or small $b$). It follows that the factor
  \begin{equation}
  q := \ee{- \alpha}
  \label{eq:noq}
  \end{equation} 
will be extremely small in the high-temperature regime (recall that
$\alpha$ is a function of $t$), and this leads us to expand $N_n$ in $q$
as
  \begin{equation}\label{eq:eqnQaad}
  N_n = \frac{ q \ee{- b e_n} }{ 1 - \eta q \ee{- b e_n} }
  = \eta^{-1} \sum\limits_{k = 1}^{\infty} (\eta q)^k \ee{- k b e_n} \, ,
  \end{equation}
with $\eta$ given in (\ref{eq:fbindex}), which is valid for any positive
$\alpha$. Thus we find
  \begin{equation}\label{eq:eqnQaae}
  \eta N = \eta \! \sum\limits_{n = 1}^{\infty} N_n \! = \!
  \sum\limits_{k = 1}^{\infty} (\eta q)^k \sum\limits_{n = 1}^{\infty}
  \ee{- k b e_n} \! = \! \sum\limits_{k = 1}^{\infty} (\eta q)^k \!
  \left[ - \frac{\sigma}{2} + \frac{1}{2}
  \sum\limits_{n = - \infty}^{\infty} \ee{- k b e_n} \! \right]
  \! ,
  \end{equation}
where we have introduced the constants $\sigma^+ = 0 \, ,$
$\sigma^- = 1$ corresponding to the half wells $W^\pm$, and
extended the meaning of the notation $e_n$ [see (\ref{eq:unitenergy})]
to negative $n$ as well. Applying the Poisson summation formula
  \begin{equation}\label{eq:eqnQaal}
  \sum\limits_{n = - \infty}^{\infty} y(n) =
  \sum\limits_{m = - \infty}^{\infty}
  \int\limits_{-\infty}^{\infty} \qd u \, y(u) \, \ee{ 2 \pi i m u } \, ,
  \end{equation}
we obtain
  \begin{equation}\label{eq:eqnQaag}
  \eta N = \sum\limits_{k = 1}^{\infty} (\eta q)^k \left[ - \frac{\sigma}{2}
  + \sqrt{ \frac{\pi}{ 4 k b } } \sum\limits_{m = - \infty}^{\infty}
  (2\sigma^{} - 1)^m \ee{ - \frac{\pi^2}{ k b } m^2 } \right],
  \end{equation}
where we note that the factor $2\sigma^{\pm} - 1$ is simply $\mp 1$.
Similarly, for the force $f$, one can find
  \begin{eqnarray}
  f & = & \sum\limits_{n = 1}^{\infty} N_n e_n \nonumber \\
  & = & \eta^{-1} \sum\limits_{k = 1}^{\infty} (\eta q)^k
  \sqrt{ \frac{\pi}{ 16 k^3 b^3 } } \sum\limits_{m = - \infty}^{\infty}
  (2\sigma^{} - 1)^m \left( 1 - \frac{2 \pi^2}{k b} m^2 \right)
  \ee{ - \frac{\pi^2}{ k b } m^2 } \, .
  \label{eq:eqnQaah}
  \end{eqnarray}

In the high-temperature asymptotic limit $b \to 0$, we have $q \to 0$ and
hence it suffices to consider only the first few terms in the sums over
$k$ [both in (\ref{eq:eqnQaag}) and (\ref{eq:eqnQaah})], and within each
term to keep only the $m = 0$ term in the sums over $m$ (the $m \ne 0$
terms being exponentially suppressed). Now, the leading $k = 1$ term in
(\ref{eq:eqnQaag}) gives
  \begin{equation}\label{eq:eqnncx}
  q = 2 N \left( \frac{b}{\pi} \right)^{1/2} + \ordo{b} \, .
  \end{equation}
This result shows that, for high temperatures, $\alpha$ tends to infinity
logarithmically. Since this leading behaviour of $q$ is independent of
$\sigma$, inserting it into (\ref{eq:eqnQaah}) gives that the leading
$\ordo{ b^{-1} } $ term of $f$ (coming from $k = 1$, $m = 0$) is also
$\sigma$-independent. Hence, the contribution coming from this term will
cancel out between the two forces $f^+$ and $f^-$ in the net force
(\ref{eq:Qnfp}).

The first nonvanishing contribution for the net force comes from the
first subleading term in $q$. Incorporating the $k = 2$ term as
well for $q$, we find
  \begin{equation}
  q = 2 N \left( \frac{b}{\pi} \right)^{1/2} + 2 N \left[ \sigma -
  \eta \sqrt{2} N \right] \frac{b}{\pi} + \ordo{ b^{3/2} } \, .
  \label{eq:eqnQaai}
  \end{equation}
Plugging this into (\ref{eq:eqnQaah}) and calculating the net force
(\ref{eq:Qnfp}) we obtain
  \begin{equation}\label{eq:eqnQaaj}
  \Delta f = \frac{N}{2} \left( \frac{t}{\pi} \right)^{1/2} + \ordo{ t^{0} } ,
  \end{equation}
which shows that the net force diverges asymptotically as $t \to \infty$
according to the square root of $t$, and that it is proportional to the
particle number $N$. Note that these are true both for the bosonic case
and the fermionic case. These results explain the high temperature
asymptotic behaviours observed by the numerical analysis in
figure\ref{fig:3}.

By incorporating higher orders, the approximation may be improved easily.
For example, by taking account of the next order, one finds
  \begin{equation}\label{eq:eqnQdaq}
  \Delta f = \frac{N}{2} \left( \frac{t}{\pi} \right)^{1/2} -
  \frac{N}{\pi} \left[ (\sqrt{2} - 1) \eta N - \frac{1}{2} \right] +
  \ordo{ t^{-1/2} } \, .
  \end{equation}
This improvement, however, is not sufficient for describing the behaviour
of the quantum force in lower (medium) temperature regimes where the
force takes its minimum.

\section{Medium temperature regime}
\label{s4}

Apart from the strength of the force, the overall shape of the force
curve -- for all temperature regimes -- agrees for bosons and fermions.
Indeed, for both of the statistical cases, the force starts with a finite
value at $t = 0$ followed by a plateau and then by a steady decrease for
larger $t$, and it exhibits even quantitatively the same high-temperature
behaviour for $t \to \infty$ as discussed in section 3. Moreover, the
force admits only one minimum between the two limiting domains (see
figure~\ref{fig:3}) for the two cases. A closer inspection of the curve
reveals, however, that in the medium temperature regime the two cases
differ in scalings with respect to the particle number $N$. Explicitly,
our numerical results (see figure~\ref{fig:4}) show that, for $\, N \gg 1
\, ,$ the value of the minimum $\Delta f_{\hbox{\scriptsize min}}$ and
the temperature $t_{\hbox{\scriptsize min}}$ where the minimum occurs are
both proportional to $N$ for bosons, while they are proportional to $N^2$
for fermions. We now give an account of these distinctive features
characterizing the statistics of the particles by analytic means.

\subsection{Bosons}
\label{ss4a}

To analyze the regime of temperatures including the minimum
$t_{\hbox{\scriptsize min}}$ for the bosonic case, we note that
interpolating between the slow logarithmic increase of $\alpha$ for high
temperature and the low-temperature behaviour (\ref{eq:eqndaw}) of
$\alpha$ (which will be derived in section~\ref{s5}) suggests the existence
of a temperature regime fulfilling
  \begin{equation}
  b \ll 1, \qquad |\alpha| \ll 1.
  \label{eq:mtr}
  \end{equation}
This is the `medium temperature regime' we wish to consider here.
Our procedure to derive the force difference is based on an approximation
for $\alpha$ fulfilling these conditions in (\ref{eq:mtr}), and we confirm the
consistency of our argument by examining the validity of the conditions
later.

\begin{figure}
\centering
\resizebox{.36\textwidth}{!}{\includegraphics{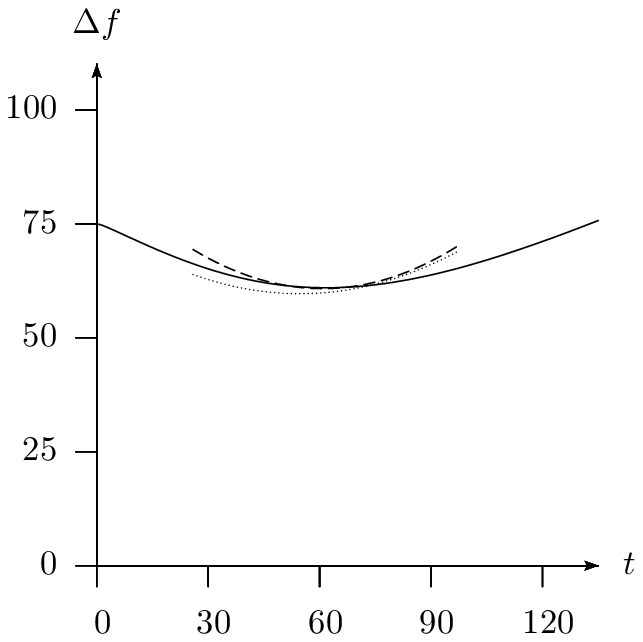}}
\hskip .1\textwidth
\raisebox{.8ex}{
\resizebox{.41\textwidth}{!}{\includegraphics{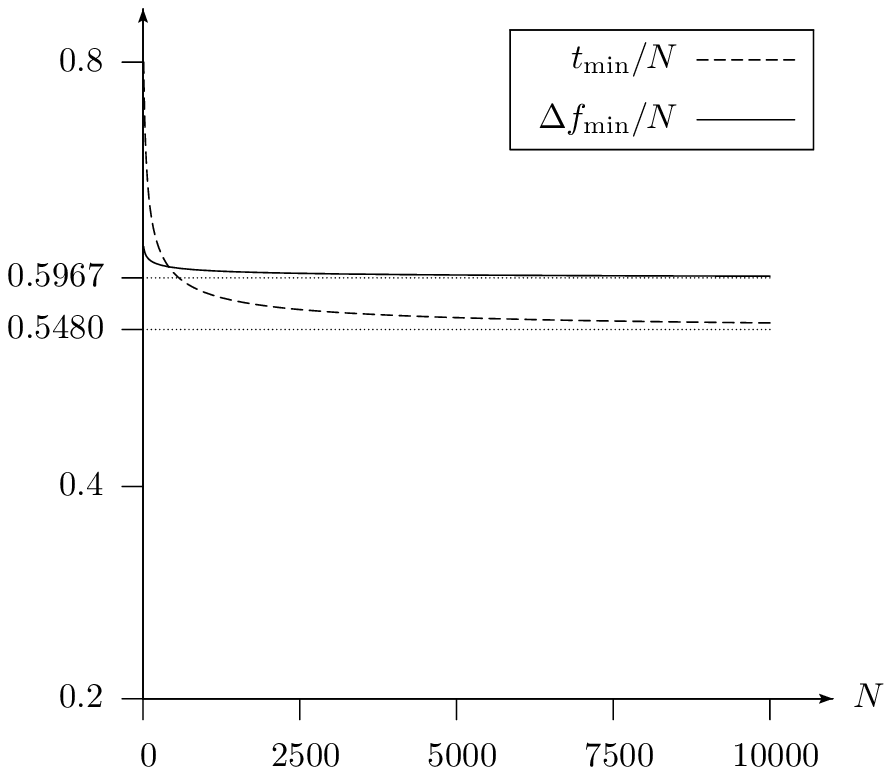}}
}
\caption{
(Left) The minimum of the force curve, for bosons, at $ N = 100 $. Solid
line: numerical computation; dashed line: the approximation
(\ref{eq:eqnQnap}); dotted line: the approximation (\ref{eq:eqnQnau}).
(Right) Numerically observed scaling behaviours of the minimum force
$\Delta f_{\hbox{\scriptsize min}}$ and the temperature
$t_{\hbox{\scriptsize min}}$ at which the minimum occurs. The limiting
values determined in (\ref{eq:eqnQnar}) are also displayed.
}\label{fig:4}
\vskip 3.4ex
\end{figure}

\begin{figure}
\centering
\resizebox{.5\textwidth}{!}{\includegraphics{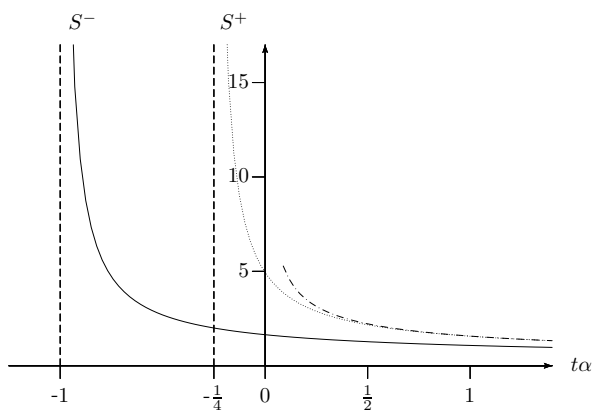}}
\caption{
The functions $S^+$ (dotted line) and $S^-$ (solid line) as defined
in (\ref{eq:eqnQnba}). They are smooth at zero, and diverge at $-e_1^+$
resp. $-e_1^-$, where the $n=1$ term in (\ref{eq:eqnQnba}) diverges. 
The dashed line depicts the approximation for $S^+$ around 1 that is
based on (\ref{eq:eqndaz}).
}\label{fig:5}
\vskip 3.4ex
\end{figure}

To proceed, we first solve the constraint (\ref{eq:constr}) for $\alpha$
by approximating the sum as
  \begin{equation}\label{eq:eqnQnay}
  N = \sum_{n=1}^{\infty} \frac{1}{ \ee{ \alpha + b e_n } - 1 } \approx
  \sum_{n=1}^{\infty} \frac{1}{ \alpha + b e_n }  \, .
  \end{equation}
This approximation must be good at least for the lower levels --- the
ones that provide the dominant contribution in the sum (note that $N_n$
falls rapidly with $n$). Let us write this in the rearranged form
  \begin{equation}\label{eq:eqnQnaz}
  \frac{N}{t} \approx \sum_{n=1}^{\infty} \frac{1}{ t \alpha + e_n } \, .
  \end{equation}
If one solves this in favor of $t \alpha$, then the solution will be a
function of $\frac{t}{N}$. Using the standard expansion formula (see,
{\it e.g.}, page 42 of \cite{GR}), we find that the summation in
(\ref{eq:eqnQnaz}) is evaluated as
  \begin{equation}\label{eq:eqnQnba}
  S(t \alpha) := \sum_{n=1}^{\infty} \frac{1}{ t \alpha + (n - \tau)^2 } =
  \left\{ \matrix{
  \frac{ \pi \tanh \pi \sqrt{t \alpha} }{2 \sqrt{t \alpha}}
  & \hbox{for}\quad W^+ \, , \cr
  \frac{\pi \sqrt{t \alpha} \coth \pi \sqrt{t \alpha} - 1}{2 t \alpha}
  & \hbox{for}\quad W^- , \cr
  } \right.
  \end{equation}
where we have used 
  \begin{equation}
  \tau := \left\{ \begin{array}{cl}
   \frac{1}{2} & \mbox{ for } W^+,  \\  0 & \mbox{ for } W^-.
  \end{array} \right.
  \label{eq:deftau}
  \end{equation}
Both formulas in (\ref{eq:eqnQnba}) are valid for negative $\alpha$ as
well, and are a smooth function of $t \alpha$ at zero, see
figure~\ref{fig:5}. Then, for a given $t$, or given $t/N$, we should solve
the transcendental equations
  \begin{equation}\label{eq:eqndbd}
  S^\pm(t \alpha^\pm) = \frac{N}{t}
  \end{equation}
to obtain $t \alpha^+$ and $t \alpha^-$ which are the values
corresponding to the half wells $W^+$ and $W^-$, respectively.

At this stage, it is informative to determine those temperatures at which
either $\alpha^+$ or $\alpha^-$ vanishes. Let $t^+_0$ and $t^-_0$ be
the temperatures such that $\alpha^+(t^+_0) = \alpha^-(t^-_0) = 0$ hold.
Using (\ref{eq:eqnQnba}), one can immediately determine $t^\pm_0$ as
  \begin{equation}
  t_0^+ = \frac{2 N}{\pi^2}, \qquad t_0^- = \frac{6N}{\pi^2}.
  \label{eq:nopi}
  \end{equation}
One then finds, numerically,
  \begin{equation}
  t^+_0 \alpha^-(t^+_0)= -0.7627, \qquad
  t^-_0 \alpha^+(t^-_0) =  0.9026 \, . 
  \label{eq:eqndba}
  \end{equation}
Our assumptions $ b \ll 1 $ and $ |\alpha^\pm| \ll 1 $ are in fact
valid at these temperatures, since we have $ t_0^\pm \sim N \gg 1 $ and
$\alpha^\pm_0 \sim 1/t = b \sim 1/N \ll 1$. Obviously, the same will
apply for the temperatures between and in the neighborhood of these two
values $ t_0^\pm$. Moreover, we can expect from these results that,
for $t \sim N$ in general,
  \begin{equation}\label{eq:eqnncy}
  \alpha^\pm \sim b \sim 1/N \ll 1 \, .
  \end{equation}
In what follows, we shall be interested in this temperature regime, where
the minimum of the force difference $\Delta f_{\hbox{\scriptsize min}}$
is found in the numerical result (see figure~\ref{fig:4}). One can show ---
see Appendix~\ref{appb} for the details --- that, for $t \sim N \, ,$ the
relative error of our approximation (\ref{eq:eqnQnay}) decreases with
increasing $N$.

Next, we derive an approximated analytical expression for the net force
$\Delta f$ obtained under (\ref{eq:eqnncy}). We do this with the help of
an integral approximation for the force sums. Presenting all the
technical details of the calculation in Appendix~\ref{appb}, here we
summarize only the result,
  \begin{equation}\label{eq:eqnQnbo}
  \frac{\Delta f}{N} \approx - \frac{1}{2} \frac{t}{N} - \Delta (t \alpha) \, .
  \end{equation}
This formula has a relative error that tends to vanish in the large-$N$
limit.

We observe from (\ref{eq:eqnQnbo}) that, in this medium temperature
domain, $\Delta f/N$ is also a function of $t/N$ as $t \alpha$ is.
Consequently, the force curve $\Delta f(t)$ is scale invariant in $N$
({\it i.e.}, it preserves its shape under the rescaling of $N$) as both
the effective range of $t$ and $\Delta f$ scale linearly with $N$. As a
special case, we see that the temperature of the minimum force occurs at
$ t_{\hbox{\scriptsize min}} \sim N $, $ \Delta f_{\hbox{\scriptsize
min}} \sim N $, confirming our observation made in the numerical analysis
in figure~\ref{fig:4}.

Let us now determine the ratios $ t / N $ and $ \Delta f / N$ in the
large-$N$ limit. One possible way to do this is to use various values of
$t/N$ to solve numerically (\ref{eq:eqndbd}) for $\alpha$ and insert the
outcomes into (\ref{eq:eqnQnbo}), and thereby reproduce the force curve
at those values. For example, the minimum of $\Delta f/N$ as the function
of $t/N$ can be estimated in this way, yielding the ratios
  \begin{equation}\label{eq:eqnQnar}
  \frac{t_{\hbox{\scriptsize min}}}{N} \approx 0.5480 \, , \hskip 10 ex
  \frac{\Delta f_{\hbox{\scriptsize min}}}{N} \approx 0.5967 \, .
  \end{equation}
These values agree with the numerical results pretty well as readily
confirmed in figure~\ref{fig:4}.

To find an analytical formula for $\Delta f$ as well, one may use some
expansion approximation of the functions $S^\pm$ [see (\ref{eq:eqnQnba})]
around some chosen value.  As an example, let us choose $t$ to be in the
vicinity of $t_0^-$. There, $ |t \alpha^-| \ll 1 \, , $ and the
equation we wish to solve becomes
  \begin{eqnarray}
  \frac{N}{t} & = & \frac{ \pi \sqrt{t \alpha^-}
  \coth \pi \sqrt{t \alpha^-} - 1 } {2 t \alpha^-} \nonumber \\
  & = & \frac{\pi^2}{6} - \frac{\pi^4}{90} t \alpha^- +
  \frac{\pi^6}{945} (t \alpha^-)^2 + \ordo{ (t \alpha^-)^3 } \, .
  \label{eq:eqnQnbf}
  \end{eqnarray}
Inverting this, one can obtain the solution as
  \begin{equation}
  t \alpha^- = \frac{5}{2} \left( \frac{t}{N} - \frac{6}{\pi^2} \right)
  + \frac{5 \pi^2}{28} \left( \frac{t}{N} - \frac{6}{\pi^2} \right)^2
  + \ordo{ \left( \frac{t}{N} - \frac{6}{\pi^2} \right)^3 } \, .
  \label{eq:eqnQnao}
  \end{equation}
As for $t \alpha^+$, we observe that it is close to the value $0.9$ [see
(\ref{eq:eqndba})], where $\, \pi \sqrt{t \alpha^+} \approx 3 \,$ and $\,
\tanh \pi \sqrt{t \alpha^+} \,$ already almost saturates to its
large-variable asymptotic value, $1$. Taking this simple asymptotic
approximation,
  \begin{equation}\label{eq:eqndaz}
  \tanh \pi \sqrt{t \alpha^+} \approx 1 \, ,
  \end{equation}
one can determine $t \alpha^+$ by the solution of
$\, N/t = \pi / (2 \sqrt{t \alpha^+} \, ) \,$, that is,
  \begin{equation}\label{eq:eqnQnbp}
  t \alpha^+ = \frac{\pi^2}{4} \left( \frac{t}{N} \right)^2 \, .
  \end{equation}
Applying the solutions (\ref{eq:eqnQnao}) and (\ref{eq:eqnQnbp}) in
(\ref{eq:eqnQnbo}) yields
  \begin{equation}\label{eq:eqnQnap}
  \frac{\Delta f}{N} \approx \frac{\pi^2}{14}
  \left( \frac{t}{N} - \frac{6}{\pi^2} \right)^2 + \frac{6}{\pi^2} \, .
  \end{equation}
This quadratic formula (\ref{eq:eqnQnap}) indicates that the
location of the minimum occurs at
  \begin{equation}\label{eq:eqnQnaq}
  \frac{t_{\hbox{\scriptsize min}}}{N} \approx
  \frac{\Delta f_{\hbox{\scriptsize min}}}{N} \approx
  \frac{6}{\pi^2} = 0.6079\ldots \, .
  \end{equation}
Figure~\ref{fig:4} shows that around the minimum the formula
(\ref{eq:eqnQnap}) reproduces the numerical curve of the force reasonably
well.

To improve our formula to achieve a better agreement between
(\ref{eq:eqnQnaq}) and (\ref{eq:eqnQnar}), we consider the approximation:
  \begin{equation}\label{eq:eqnQnas}
  \tanh x \approx \frac{ \tanh x^* + (x - x^*) }{ 1 +
  \tanh x^* \cdot (x - x^*) } \, .
  \end{equation}
This formula is precise up to the quadratic Taylor term at $x^*$, and
behaving much better than the quadratic Taylor polynomial approximation
in a larger neighborhood. For the expansion point $\, x^* \equiv \pi
\sqrt{ (t \alpha^+)^* } \, ,$ we may simply choose the above-mentioned
value, 3. Using the notation $\, t^*/N := S^+\left( (t\alpha^+)^* \right)
\,$ --- which is close to $t_0^-/N$ --- and assuming an expansion of the
form
  \begin{equation}\label{eq:eqnQnat}
  t \alpha^+ - (t \alpha^+)^* = c_1 \left(\frac{t}{N} - \frac{t^*}{N}\right) +
  c_2 \left(\frac{t}{N} - \frac{t^*}{N}\right)^2 + \ordo{ \left(\frac{t}{N} -
  \frac{t^*}{N}\right)^3 } \, ,
  \end{equation}
we obtain a quadratic approximation for $t \alpha^+$ that is better
than (\ref{eq:eqnQnbp}).  Accordingly, the force difference is obtained in
the improved form,
  \begin{equation}\label{eq:eqnQnau}
  \frac{\Delta f}{N} \approx
  0.5121 \cdot \left( \frac{t}{N} - 0.5465 \right)^2 + 0.5967 \, ,
  \end{equation}
having its minimum at
  \begin{equation}\label{eq:eqnncz}
  \frac{t_{\hbox{\scriptsize min}}}{N} \approx 0.5465 \, , \hskip 10 ex
  \frac{\Delta f_{\hbox{\scriptsize min}}}{N} \approx 0.5967 \, .
  \end{equation}

We remark that the parabolic approximation used above is valid only
locally and not suitable for describing the whole medium temperature
domain. To find a formula valid for other values of temperatures, we may
simply expand at the value of the interest and/or using approximate
formulas for the other parts of the functions $S$. A more universally
valid approximate formula, which describes the force curve in the whole
medium temperature region, is obviously desirable but it is rather
difficult to find at the moment.

\subsection{Fermions}
\label{ss4b}

The numerical results shown in figure~\ref{fig:4} suggest that, for fermions, the
medium temperature regime that includes the temperature of the minimum
force may be defined as the domain where $t \sim N^2 \, ,$ $\Delta f \sim
N^2$ are fulfilled. Now we investigate this regime to seek an analytic
approximation of the force curve there.

As we did for the bosonic case, we first determine $\alpha$.  Our
integral approximation presented in Appendix~\ref{appb} gives
  \begin{equation}\label{eq:nbw}
  N = \sum_{n=1}^{\infty} \frac{1}{ \ee{\alpha + y_n^2} + 1 } \approx
  \frac{\tau - 1/2}{ \ee{\alpha} + 1 } + \frac{1}{\sqrt{b}} \; I(\alpha)
  \end{equation}
with  the Fermi-Dirac integral $I(\alpha)$ given by 
  \begin{equation}
  I(\alpha) := \int_0^\infty \frac{ \qd y }{ \ee{\alpha + y^2} + 1 } \, .
  \label{eq:noI}
  \end{equation}
Since $\, \frac{1}{ \ee{\alpha} + 1 } \approx N_1 < 1 \, ,$ for large $N$
we can simplify (\ref{eq:nbw}) to
  \begin{equation}\label{eq:nby}
  N \approx \sqrt{t} \,\, I(\alpha)  \hskip 4 ex \hbox{or} \hskip 4 ex
  t \approx \frac{N^2}{I(\alpha)^2} \, ,
  \end{equation}
where, for temporary convenience, we solve our condition for $t$ as a
function of $\alpha$, instead of the reverse. We observe that $\, t
\sim N^2 \,$ implies $ \alpha = \ordo{N^0}$, unlike in the medium
temperature regime of the bosonic case.

Since there arises no difference between $\alpha^+$ and $\alpha^-$ in
this leading order, we need to consider the subleading term for $\Delta
\alpha.$ This can be derived from (\ref{eq:nbw}) as
  \begin{equation}\label{eq:nbz}
  0 = \Delta N \hskip -.2ex \approx \Delta \hskip -.4ex
  \left[ \frac{\tau - \frac{1}{2}}{ \ee{\alpha} +
  1 }\right] + \frac{1}{\sqrt{b}} \, I'(\alpha) \Delta \alpha
  \hskip .05ex , \hskip 3ex \Delta \alpha \approx \frac{1}{2N}
  \frac{ I(\alpha) }{ \left( \ee{\alpha} + 1 \right) I'(\alpha) } \, ,
  \end{equation}
where we have used the fact that, in leading order, the difference
between $\alpha^+$ or $\alpha^-$ can be ignored (and we have again
eliminated $t$ as a function of $\alpha$).

{}For the force, our integral approximation provides
  \begin{equation} 
  f = \sum_{n=1}^{\infty} \frac{e_n}{ \ee{\alpha + y_n^2} + 1 } = b^{-3/2}
  \sum_{n=1}^{\infty} \frac{y_n^2 \Delta y}{ \ee{\alpha + y_n^2} + 1 }
  \approx {\cal O}\big(t^0\big) + t^{3/2} \! \int_0^\infty
  \frac{ y^2 \qd y }{ \ee{\alpha + y^2} + 1 } , \hskip 1ex
  \label{eq:nof}
  \end{equation}
and hence by integration by parts and (\ref{eq:nbz}) we obtain
  \begin{equation}
  \Delta f \approx t^{3/2} \int_0^\infty \frac{ - \ee{\alpha + y^2}
  \Delta \alpha }{ \left( \ee{\alpha + y^2} + 1 \right)^2 } \, y^2 \qd y
  = \frac{- t^{3/2} }{2} \Delta \alpha \int_0^\infty
  \frac{\qd y}{ \ee{\alpha + y^2} + 1 } \approx \frac{N^2}{4} J(\alpha),
  \label{eq:ncc}
  \end{equation}
with
  \begin{equation}\label{eq:ncd}
  J(\alpha) := \frac{-1}{ \left( \ee{\alpha} + 1 \right) I(\alpha)
  I'(\alpha) } = \frac{-2}{(\ee{\alpha} + 1) \,
  \qd I^2(\alpha) / \qd \alpha} \, .
  \end{equation}
All these approximations will improve for larger values of $t$ and $N$.
We can see in (\ref{eq:ncc}) that, having $ \alpha = \ordo{N^0}$, not
only $t$ but also $\Delta f$ scale with $N^2$ in the medium temperature
regime, and consequently the force curve is again scale invariant in $N$
(shape-preserving under rescaling $N$) in this regime.

As before, for a given $t$, one can obtain $\alpha$ and  $\Delta f$ by
solving a transcendental equation numerically or by using some analytical
approximate formula. For instance, determining the location of the
minimum --- in essence, the minimum of $J(\alpha)$ --- via numerical
solution, one finds
  \begin{equation}\label{eq:ncm}
  \alpha_{\hbox{\scriptsize min}} = -2.567 , \hskip .8ex
  J(\alpha_{\hbox{\scriptsize min}}) \approx 1.813 ,
  \hskip .8ex t_{\hbox{\scriptsize min}} \approx 0.444 N^2
  \hskip -.2ex ,
  \hskip .8ex \Delta f_{\hbox{\scriptsize min}} \approx 0.453 N^2
  \hskip -.2ex .
  \end{equation}
These values are in apparent accord with the fully numerical results in
figure~\ref{fig:4}.

For an analytical approach, first one can take the standard asymptotic
series for the Fermi-Dirac integral $I(\alpha)$. According to
\cite{Stoner}, in the interval $\, \alpha \in [ -3.696, -1.314 ] \,$ the
truncation
  \begin{equation}\label{eq:ncf}
  I(\alpha) \approx \sqrt{-\alpha}\left[ 1 - \frac{\pi^2}{24}
  \frac{1}{(-\alpha)^2} \right]
  \end{equation}
of the asymptotic series is the best available approximation for
$I(\alpha)$. Unfortunately, this is not sufficient for our purpose
because the approximated $J(\alpha)$ obtained from this does not possess
a minimum. However, noticing that, for $I'(\alpha)$, a better
approximation is obtained by the further truncated
  \begin{equation}\label{eq:ncg}
  I'(\alpha) \approx \frac{1}{2 \sqrt{-\alpha}}
  \end{equation}
we can find, using this in $J(\alpha)$, a minimum in the temperature
regime we are considering.  By numerically solving the arising
transcendental equation, one finds
  \begin{equation}\label{eq:nch}
  \alpha_{\hbox{\scriptsize min}} \approx -1.95 , \hskip 1.5ex
  J(\alpha_{\hbox{\scriptsize min}}) \approx 1.96 , \hskip 1.5ex
  t_{\hbox{\scriptsize min}} \approx 0.611 N^2 ,
  \hskip 1.5ex \Delta f_{\hbox{\scriptsize min}} \approx 0.49 N^2 ,
  \end{equation}
which are not quite precise compared to (\ref{eq:ncm}).

\begin{figure}
\centering
\resizebox{.39\textwidth}{!}{\includegraphics{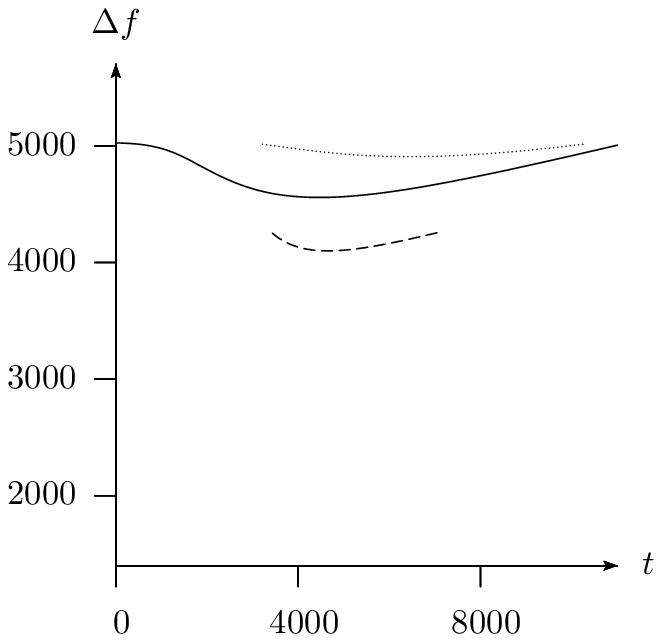}}
\hskip .1\textwidth
\raisebox{.6ex}{\resizebox
{.41\textwidth}{!}{\includegraphics{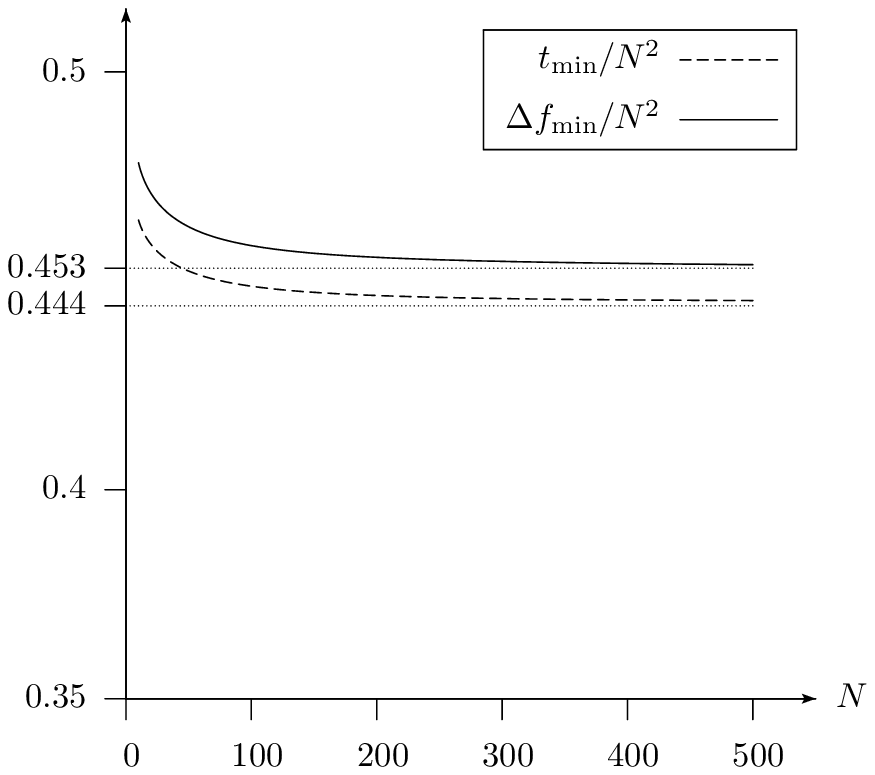}}}
\caption{
(Left) The minimum of the force curve for fermions, at $ N = 100 \, . $
Solid line: numerical computation; dotted line: the approximation
corresponding to (\ref{eq:nch}); dashed line: the approximation
corresponding to (\ref{eq:ncq}). (Right) the scaling behaviour of
$t_{\hbox{\scriptsize min}}$ and $\Delta f_{\hbox{\scriptsize min}}$.
The constants determined at (\ref{eq:ncm}) are also displayed.
}\label{fig:6}
\vskip 3.4ex
\end{figure}

To find a second, better approximation of $I(\alpha)$, we utilize the
fact that its integrand, $\, \frac{1}{ \ee{\alpha + y^2} + 1 } \,$ is
close to a tangent hyperbolic function (reflected and shifted) on $\, y
\in [0, \infty) \, ,$ for negative $\alpha$. Specifically, we may write
  \begin{eqnarray}
  I(\alpha) & \approx & \int_0^\infty \frac{p}{2} \left[ 1 - \tanh
  \sqrt{-\alpha} \left( y - \sqrt{-\alpha} \right) \right] \qd y
  \nonumber \\  & = & \frac{p}{2\sqrt{-\alpha}} 
  \left[ -2\alpha + \ln (1 + \ee{2\alpha}) \right]
  \label{eq:ncn}
  \end{eqnarray}
with $\, p := \frac{ 1 + \ee{2\alpha} }{ 1 + \ee{\alpha} } \,.$ Here, the
integrand is chosen to be simple enough but still to reproduce the true
integrand exactly at $ y = 0 $ and decreases to the half of the $ y = 0 $
value around the same point with similar steepness. Omitting $ \ordo{
\ee{4\alpha} }$ terms can reduce the obtained formula to
  \begin{equation}\label{eq:nco}
  I(\alpha) \approx \frac{p}{2\sqrt{-\alpha}} \left( -2\alpha +
  \ee{2\alpha} \right) ,  \hskip 6 ex  I^2(\alpha) \approx
  p^2 \left(-\alpha + \ee{2\alpha}\right) \, .
  \end{equation}
Since no further available simplification can render the resulting
transcendental equation analytically solvable for $\alpha$, we shall use
(\ref{eq:nco}) to expand our $J(\alpha)$ approximation around, say, $
\alpha^* = -2.5 $ to second order, and determine the minimum of the
quadratic Taylor polynomial, which polynomial we can rewrite in the form
  \begin{equation}\label{eq:ncp}
  J(\alpha) \approx 0.134 (\alpha + 2.48 )^2 + 1.64 \, .
  \end{equation}
This way we reach
  \begin{equation}\label{eq:ncq}
  \alpha_{\hbox{\scriptsize min}} \approx -2.48 , \hskip 1.3ex
  J(\alpha_{\hbox{\scriptsize min}}) \approx 1.64 , \hskip 1.3ex
  t_{\hbox{\scriptsize min}} \approx 0.466 N^2 ,
  \hskip 1.3ex \Delta f_{\hbox{\scriptsize min}} \approx 0.411 N^2 ,
  \end{equation}
which agree better with the numerical results; see (\ref{eq:ncm}) and
figure~\ref{fig:4}. Should one need a further improved formula, the
approach can be repeated with some enhanced or more diligent choice for
approximating the integrand in $I(\alpha)$.

\section{Low temperature regime}
\label{s5}

Having analyzed the quantum force on the partition in high and medium
temperature regimes, we now consider the force under low temperatures in
this section.

\subsection{Bosons}
\label{ss5a}

To discuss the bosonic case, let us first approach from the
zero-temperature end. Since at exactly zero temperature all the particles
sit on the lowest available level which is the ground state for bosons,
we immediately obtain
  \begin{equation}\label{eq:eqndar}
  f^+(0) =  e_1^+ N = \frac{N}{4} \, , \qquad  f^-(0) =  e_1^- N = N \, .
  \end{equation}
This implies
  \begin{equation}
  \Delta f (0) = \frac{3}{4} N \, ,
  \label{eq:bfzt}
  \end{equation}
which is nonzero and is proportional to $N$.  

\begin{figure}
\centering
\resizebox{.378\textwidth}{!}{\includegraphics{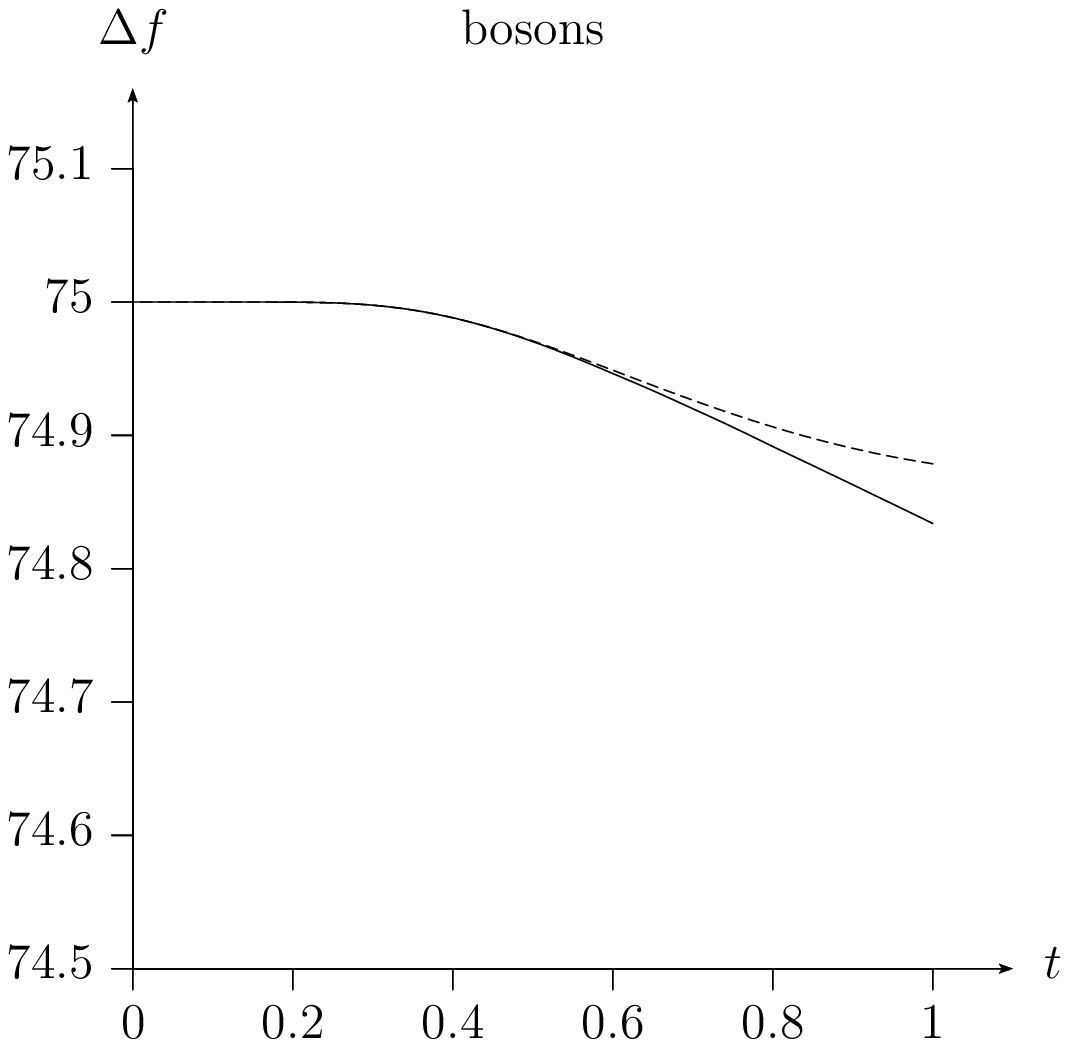}}
\hskip .1\textwidth
\resizebox{.42\textwidth}{!}{\includegraphics{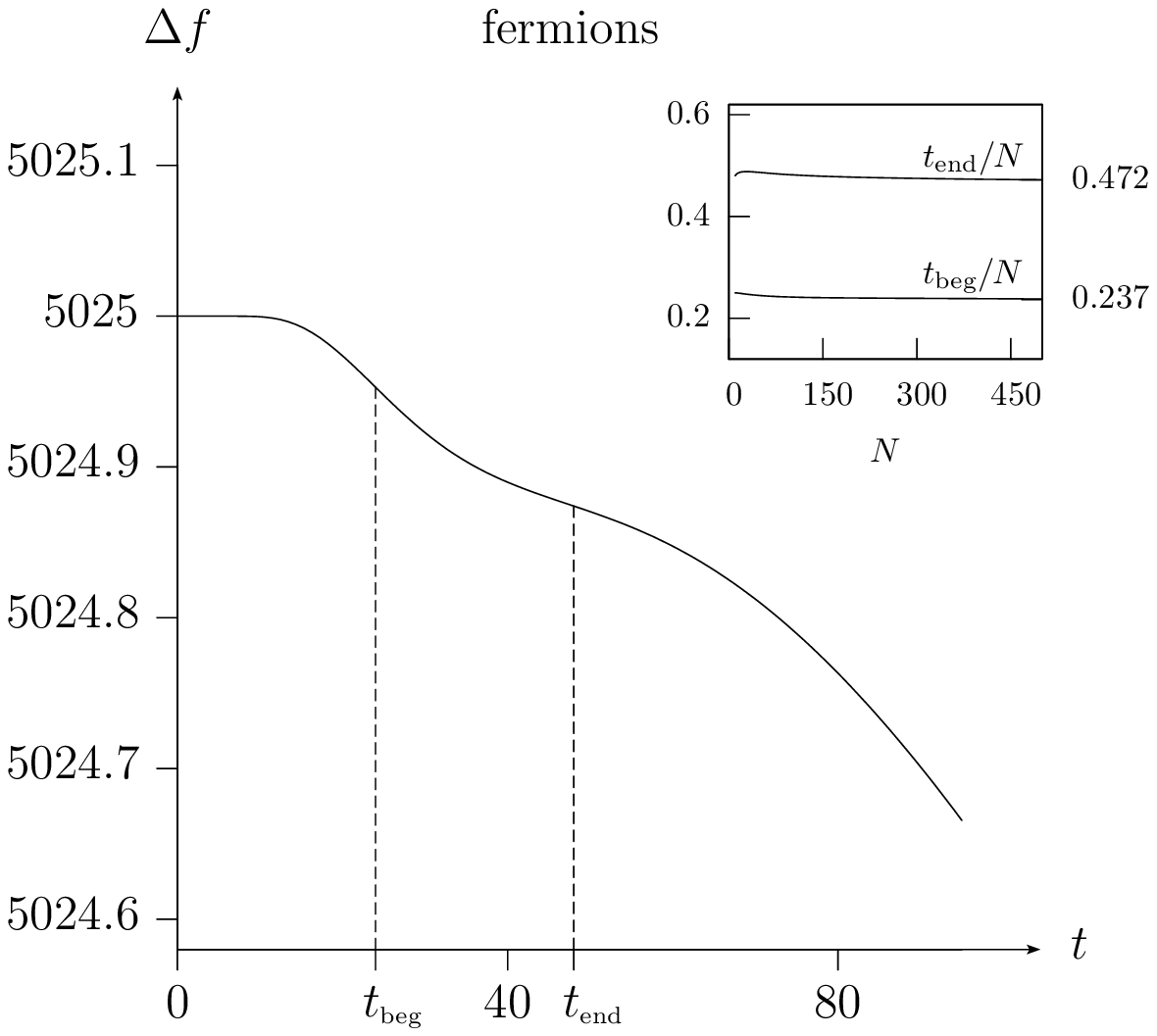}}
\caption{
The low-temperature behaviour of the net force $\Delta f$, for $\,N
= 100 \, .$
(Left) the bosonic case. Solid line: numerical computation, dashed line:
the two-level approximation (\ref{eq:eqnQnfres}).
(Right) the numerical results for the fermionic case. Observe the steplike
shape, characterized by the temperature values at the two points of
inflection. The inset displays $ \tbeg / N $ and $ \tend / N \, ,$
as the function of $N$. They appear to tend to constants that are
approximately $0.237$ and $0.472$, respectively.
Within an error, the second number is just the double of the first one.
}\label{fig:7}
\vskip 3.4ex
\end{figure}

When the temperature is slightly above zero, the particles occupy the
lower excited levels in addition to the ground state, and the transition
from the ground level to the upper levels is more extensive in the half
well $W^+$ than $W^-$, because the subsequent energy levels have a
smaller difference in $W^+$ than $W^-$ (see (\ref{eq:unitenergy})).
Consequently, the force difference will decrease as $t$ grows from zero.
Indeed, in the two-level approximation for bosons, where the higher
levels are treated as still completely unoccupied, the net force is found
to be \cite{FMT}
  \begin{equation}
  \Delta f(t) \approx \frac{3}{4} N + (3\, e^{-3/t} - 2\, e^{-2/t}) \, ,
  \label{eq:eqnQnfres}
  \end{equation}
which accounts for the decrease, see figure~\ref{fig:7}. Note that $\Delta
f(t)$ starts to decrease for $ t \sim 1 \, ,$ irrespective of the
particle number $N$. Incidentally, we mention that the low temperature
behaviour for $\alpha$ is
  \begin{equation}\label{eq:eqndaw}
  \alpha \approx - b e_1 + \ln \left( 1 + \frac{1}{N} \right) \, ,
  \end{equation}
which is a straightforward consequence of the approximation $\, N_1
\approx N \, .$

\subsection{Fermions}
\label{ss5b}

To study the zero temperature limit for fermions, we recall that
the lowest $N$ levels are occupied at $t = 0$, and from this we obtain
  \begin{equation}\label{eq:eqndas}
  f^+(0) = \! \sum\limits_{n = 1}^{N} e_n^+ =
  \mbox{\footnotesize $\displaystyle
  \frac{N (4N^2 - 1)}{12} ,
  $}
  \hskip 2.5ex
  f^-(0) = \! \sum\limits_{n = 1}^{N} e_n^-  =
  \mbox{\footnotesize $\displaystyle
  \frac{N (N + 1) (2N + 1)}{6} ,
  $}
  \end{equation}
which implies
  \begin{equation}
  \Delta f (0) =
  \mbox{\footnotesize $\displaystyle
  \frac{N(2N+1)}{4} \, .
  $}
  \label{eq:ffzt}
  \end{equation}
Observe that, in contrast to the bosonic case, the force difference in
the fermionic case is (for $N \gg 1$) proportional to $N^2$ in the
limit $t \to 0$.

\begin{figure}
\centering
\resizebox{.42\textwidth}{!}{\includegraphics{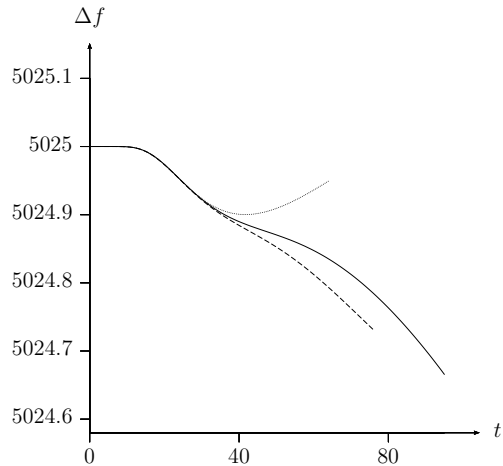}}
\caption{
The two- and semi-four-level approximations (dotted and dashed lines,
respectively) near the fermionic step (solid line) for $\, N = 100 \, .$
}\label{fig:8}
\vskip 3.4ex
\end{figure}

We have learned that the fermionic net force $\Delta f(t) $ differs from
the bosonic one quantitatively in the zero temperature limit $\Delta f(0)
$ --- it is proportional to $N^2$ for fermions while the order is $N$ for
bosons. Another quantitative difference can be observed in the first
`turning point\rq, namely the temperature where $\Delta f(t) $ starts to
decrease when the temperature $t$ is increased from zero -- the turning
point occurs at around $ t \sim N $ for fermions (as confirmed
numerically) while it occurs at around $ t \sim 1$ for bosons. Besides,
there is a qualitative difference between the fermionic and the bosonic
cases -- the fermionic curve exhibits a single, very small but
unmistakable `steplike' pattern (or a depression) for any $N$ during the
initial decrease as shown in figure~\ref{fig:8}. This step occurs at a
temperature proportional to $N$ (see figure~\ref{fig:8}), and the net force
differs there from the $\ordo{N^2}$ zero temperature value by only an
$\ordo{N^0} = \ordo{1}$ amount.

These numerically observed properties can be understood analytically as
follows. For fermions, the analogue of the bosonic two-level
approximation corresponds to the situation where only the occupation of
the $N$th and $(N + 1)$th levels ({\it i.e.}, the two levels closest to
the Fermi level) are different from the zero temperature value. On each
side, this imposes the approximate equation
  \begin{equation}\label{eq:ncr}
  N_N + N_{N+1} = 1 , \hskip 3.ex N_{N+1} =
  \frac{1}{\ee{\alpha + b e_{N+1}}
  + 1} = 1 - N_N = \frac{1}{\ee{-\alpha - b e_{N}} + 1} ,
  \end{equation}
which can be solved for $\alpha$ as
  \begin{equation}\label{eq:eqndax}
  \alpha \approx - \frac{b}{2} \left( e_N + e_{N+1} \right) \, .
  \end{equation}
Let us observe that, with this approximation for $\alpha$,
  \begin{equation}\label{eq:nct}
  N_{N+l} = \frac{1}{\ee{\alpha + b e_{N+l}} + 1} \approx 1 - N_{N+1-l}
  = \frac{1}{\ee{-\alpha - b e_{N+1-l}} + 1} \approx
  \frac{1}{\ee{(2l-1) bN} + 1}
  \end{equation}
($l = 1, 2, \ldots$) in the leading order of $N$ in the exponents.
Keeping only two nontrivial levels for calculating the forces as well
({\it i.e.,} $ l = 1 $ only), we have
  \begin{eqnarray}
  f & = & \sum_{n=1}^\infty e_n N_n = \sum_{n=1}^N e_n -
  \sum_{n=1}^N e_n (1 - N_n) + \sum_{n=N+1}^\infty e_n N_n \nonumber \\
  & \approx & f(0) + \left( e_{N+1} - e_N \right) N_{N+1} \, .
  \rule{0ex}{3.ex}
  \label{eq:ncs}
  \end{eqnarray}

At this point, we observe that a cancellation in the leading order of $N$
takes place, since we have $\, e_N \approx e_{N+1} \approx N^2 \,$ but
$\, e_{N+1} - e_N = 2 N + (1 - 2\tau) \,$ (and, in general, $\, e_{N+l} -
e_{N+1-l} = (2l-1) [2 N + (1 - 2\tau) ] \, $) with $\tau$ defined in
(\ref{eq:deftau}), which is only $ \ordo{N} $. An even higher
cancellation will occur in the force difference, as this leading
$\ordo{N}$ difference is the same on the two sides and, according to
(\ref{eq:nct}), $N_{N+1}$ is also the same on the two sides in the
leading order of $N$. To obtain a nonvanishing contribution, we use for $
N_{N+1}^\pm$ the first-order Taylor approximation
  \begin{equation}\label{eq:ncu}
  \frac{1}{ \ee{X + \Delta X} + 1 } \approx \frac{1}{\ee{X} + 1} -
  \frac{\ee{X}}{ \left( \ee{X} + 1 \right)^2} \, \Delta X
  \end{equation}
with $ X = bN $ and appropriate $\Delta X^\pm$. This gives
  \begin{eqnarray}
  \Delta f & \approx & \Delta f(0) + \frac{1}{\ee{bN} + 1} - b\left(
  N + \frac{1}{2} \right) \frac{\ee{bN}}{ \left( \ee{bN} + 1 \right)^2}
  \nonumber \\
  & \approx & \Delta f(0) + \frac{1}{\ee{bN} + 1} - bN
  \frac{\ee{bN}}{ \left( \ee{bN} + 1 \right)^2} \, .
  \label{eq:eqnday}
  \end{eqnarray}
Now we can see that this last expression contains $t$ only in the
combination $bN = N/t$, which explains why the decrease of the net force
from the zero temperature value starts at $ t \sim N \, . $ Further, it
is also visible that this decrease at $ t \sim N $ is only a small
$\, \ordo{N^0} = \ordo{1} \,$ phenomenon.

This two-level approximation is not sufficient to explain the step in the
curve pertinent to the fermionic case (see figure~\ref{fig:8}). However, we
may apply a semi-four-level approximation, that is, we assume four
nontrivially occupied levels but use the two-level-approximated $\alpha$.
The net force evaluated by an analogous procedure then becomes
  \begin{equation}\label{eq:ncv}
  \Delta f \approx \Delta f(0) + \frac{1}{\ee{bN} + 1} + \frac{3}
  {\ee{3bN} + 1} - \frac{bN \; \ee{bN}}{ \left( \ee{bN} + 1 \right)^2} -
  \frac{13 bN \; \ee{3bN}}{ \left( \ee{3bN} + 1 \right)^2} \, .
  \end{equation}
{}From this, we are able to read off $\tbeg$ and $\tend$ that
characterize the location of the step in figure~\ref{fig:8} by numerically
determining the points of inflection as a function of $t/N$. The results
  \begin{equation}\label{eq:ncw}
  \tbeg/N \approx 0.239 \, , \hskip 6 ex \tend/N \approx 0.426
  \end{equation}
are indeed close to the numerically observable values $0.237$ and
$0.472$, respectively.

However, as seen in figure~\ref{fig:8}, for the curve over the entire low
temperature regime, this semi-four-level approximation is less
satisfactory. Our further investigation shows that the net force $\Delta
f$ is extremely sensitive to the error in $\alpha$, and that
incorporating two more levels provides only a smaller contribution
compared to the change caused by the error. It seems, therefore, that a
full four-level approximation for $\alpha$ is required for the net force
to reproduce the numerical curve more precisely.

\section{The partition shift $\Delta l$ and the transfer number $\Delta N$}
\label{s6}

\begin{figure}
\centering
\resizebox{.33\textwidth}{!}{\includegraphics{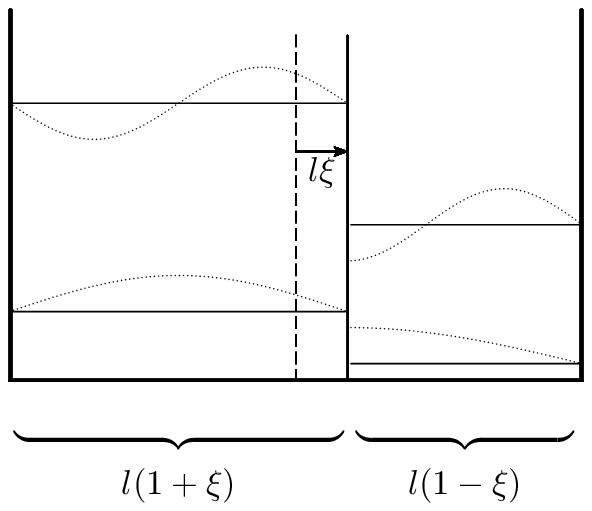}}
\hskip .1\textwidth
\resizebox{.3993\textwidth}{!}{\includegraphics{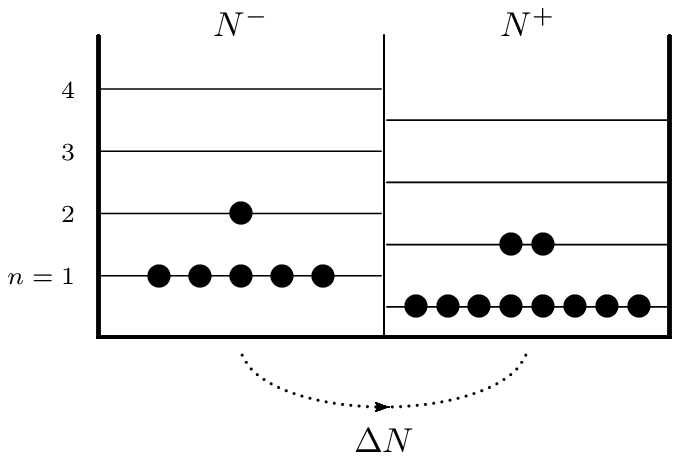}}
\caption{
(Left) the shift $\Delta l = l\xi$ in the position of the partition.
(Right) the redistribution of particles $\Delta N = N_+ - N$. Both
$\Delta l$ and $\Delta N$ are determined from the stability, $\Delta F(0)
= 0$.
}\label{fig:9}
\vskip 3.4ex
\end{figure}

We have seen that the quantum effect caused by a set of nontrivial
boundary conditions manifests in the force that arises on the partition
in a potential well. For actual observation of the effect, however, there
may be other quantities which are more readily measurable than the force
itself. In this section we mention briefly two examples of such
quantities: the shift $\Delta l$ in the position of the partition and the
transfer $\Delta N$ of particles between the two half wells (see
figure~\ref{fig:9}). For simplicity, our discussions are mostly restricted
to the zero temperature limit $t \to 0$.

To discuss the first example, suppose that the partition is allowed to
move freely in the well. Due to the pressure, the partition will then
move and acoordingly the widths of the half wells $W^\pm$ change as
  \begin{equation}
  W^+: \, l \to l(1-\xi), \qquad  W^-: \, l \to l(1+\xi),
  \label{eq:noW}
  \end{equation}
so that the net force vanishes, $\Delta F(0) = 0$.
The portion $\xi$ of the shift $\Delta l = l\xi$ is thus determined by  
  \begin{equation}
  0 = \Delta F(0) =  \frac{(\hbar\pi)^2}{m}\frac{1}{l^3}
  \left[\frac{f^-(0)}{(1+\xi)^3} - \frac{f^+(0)}{(1-\xi)^3}\right].
  \label{eq:noD}
  \end{equation}
{}From this we find
  \begin{equation}
  \xi = \frac{r - 1}{r + 1} \qquad \hbox{with} \qquad
  r = \left[\frac{f^-(0)}{f^+(0)}\right]^{1/3}.
  \label{eq:xmove}
  \end{equation}

{}For the bosonic case, (\ref{eq:bfzt}) implies
$ r = \sqrt[3]{4}$ and hence
  \begin{equation}
  \xi
  = \frac{\sqrt[3]{4} - 1}{\sqrt[3]{4} + 1}
  \approx 0.227.
  \label{eq:bshift}
  \end{equation}
This shows that the shift is rather large reaching nearly a quarter of
the original width. Notice that for the bosonic case the portion $\xi$ of
the shift is independent of the particle number $N$.

For the fermionic case, in contrast, the force limit (\ref{eq:ffzt})
implies the factor $r \approx 1 + 1/(2N)$ for $N \gg 1$ to each up and
down spin, and hence the portion $\xi$ in (\ref{eq:xmove}) becomes
  \begin{equation}
  \xi  \approx \frac{1}{4} \frac{1}{N},
  \label{eq:fshift}
  \end{equation}
which is quite small for large $N$ ({\it e.g.}, less than 1 percent even
for $N = 100$). The result (\ref{eq:fshift}) indicates that the fermionic
shift is much smaller than the bosonic one (\ref{eq:bshift}) and is
almost undetectable for large $N$. We thus learn that the spread of
particles over the levels according to the Fermi-Dirac statistics has the
effect of balancing the partition near the centre.

\begin{figure}
\centering
\resizebox{.42\textwidth}{!}{\includegraphics{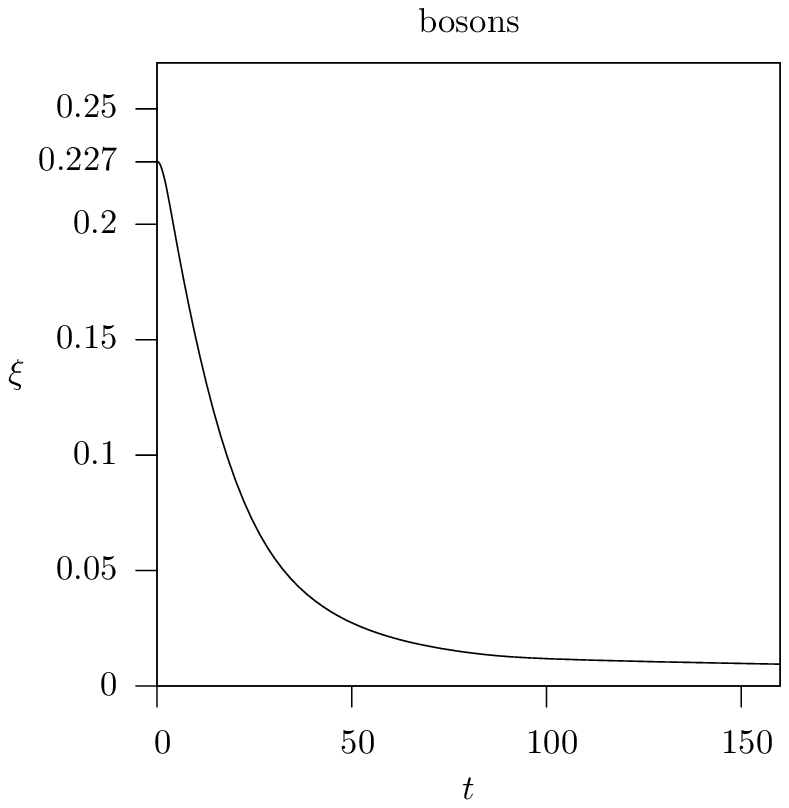}}
\hskip .1\textwidth
\resizebox{.4375\textwidth}{!}{\includegraphics{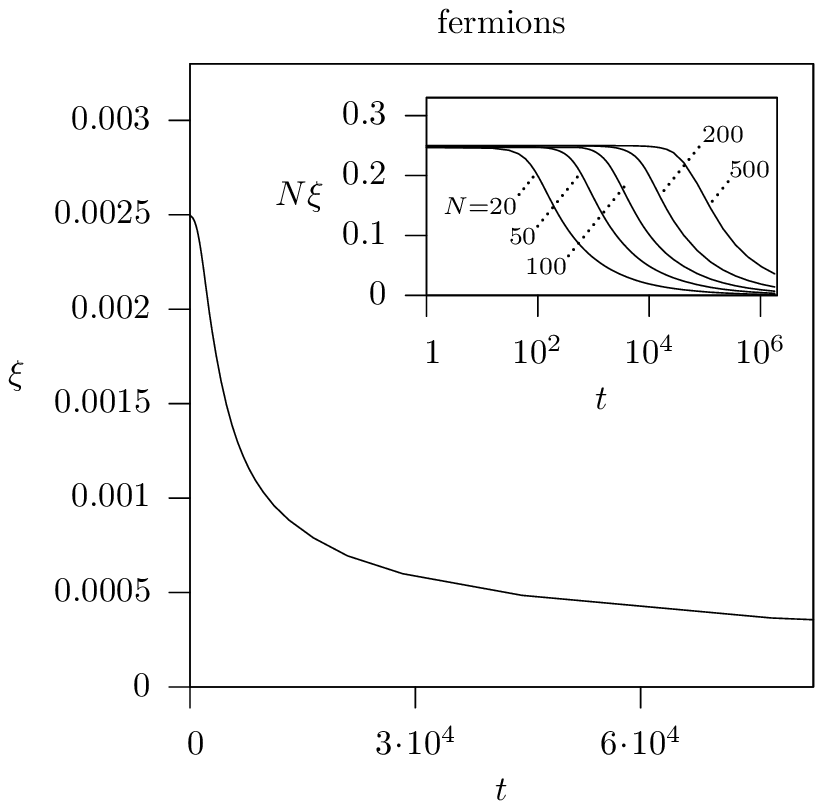}}
\caption{
The equilibrial shift of the partition as the funtion of temperature
obtained numerically for $ N = 100$; (Left) for bosons and (Right) for
fermions. The inset in the fermionic case shows the quadratic
$N$-dependence of the combination $N \xi(t)$.
}\label{fig:10}
\vskip 3.4ex
\end{figure}

When the temperature $t$ increases from zero,  the shift $\xi$ decreases
steadily for all $t$ in both bosonic and fermionic cases. This is due to
the fact that as $t$ becomes higher the forces $f^-$ and $f^+$ grow
faster than their difference, rendering the ratio $r$ closer to unity.
This is seen clearly in our numerical analysis shown in
figure~\ref{fig:10}. Interestingly, it also shows that, for fermions for
instance, the turning point of the shift $\xi$ after the initial plateau
in the vicinity of $t = 0$ has a curious scaling property under the
change of the particle number $N$. Namely, the inset of figure~\ref{fig:10}
indicates that the turning point temperature of $\xi$ depends linearly on
$N$ (since, for $N \xi$, it scales as $N^2$), which is the same scaling
law found for the force $\Delta f$ for its turning point in the low
temperature regime.

In our second example, we suppose that, unlike in the previous
situation,
the partition stays at the centre but instead we consider to
move particles from one half well to the other until the equilibrium
$\Delta F(0) = 0$ is achieved. Let $N_+$ and $N_-$ be the redistributed
number of particles in the half wells $W^\pm$, {\it i.e.},
  \begin{equation}
  W^+: \, N \to N_+, \qquad  W^-: \, N \to N_-, \qquad
  \hbox{with}\quad N_+ + N_- = 2N.
  \label{eq:noN}
  \end{equation}
The amount $\Delta N = N_+ - N$ of moved particles with respect to the
original numbers $N$ is another good measure of how far the original
situation is away from the mechanical equilibrium.  The condition to
determine this number $\Delta N$ is just $r = 1$ with $r$ given in
(\ref{eq:xmove}).

For the bosonic case, where we have (\ref{eq:eqndar}), this condition
implies $N_+/(4N_-) = 1$ and, accordingly, the stability is achieved
by the ratio
  \begin{equation}
  \frac{N_+}{N_-} = 4 \, .
  \label{eq:no4}
  \end{equation}
Thus the redistributed numbers read
  \begin{equation}
  N_+ = \frac{8}{5}N \qquad \hbox{and}\qquad N_- = \frac{2}{5}N,
  \label{eq:no8}
  \end{equation}
which implies that a large portion (60\%) of particles must be moved from
one well to the other to achieve the equilibrium. As in the previous
example, this portion does not depend on the particle number $N$. This
second example also illustrates the fact that the bosonic case is
macroscopically far from the mechanical equilibrium and hence the effect
may easily be observed experimentally.

In contrast, the fermionic case has the limit (\ref{eq:ffzt}), and the
equilibrium condition $r= 1$ implies $(N_+/N_-)^3\cdot (1 +
3/(2N_-))^{-1} = 1$. This is achieved with the ratio,
  \begin{equation}
  \frac{N_+}{N_-} \approx 1 + \frac{1}{2}\frac{1}{N}.
  \label{eq:noo}
  \end{equation} 
The outcome shows that the instability caused by the net force on the
partition is extremely small so that even the redistribution by a single
particle from $W^-$ to $W^+$ can reverse the instability ($\Delta F(0) <
0$). We thus learn that, in terms of redistribution, the effect of the
quantum force on the partition is almost invisible for the fermionic
case. Again, we see that the fermionic setting is macroscopically close
to the mechanical equilibrium.

\section{Conclusion and Discussions}
\label{s7}

In the present paper, we have investigated physical consequences of
nontrivial boundary conditions in a quantum well realized by a partition
placed at the centre of the well. We have examined the pressure appearing
on the partition due to the distinct boundary conditions on the two sides
(Dirichlet on the left and Neumann on the right), taking into account of
the thermal effects under finite temperatures as well as the different
statistics of the particles in the well. We have found that, for both
bosons and fermions, the net force $\Delta f(t)$ acting on the partition
in the quantum well is nonzero at the zero temperature limit $t \to 0$,
and remains practically constant for extremely low temperatures before it
starts to decrease gradually for higher (but still low) temperatures.
Knowing that the energy spectrum is different in the two half wells $W^-$
and $W^+$ separated by the partition for all energy regimes, this
property is not unexpected. What is unexpected, however, is that this
decrease stops at a certain temperature $t_{\hbox{\scriptsize min}}$ and
afterwards the net force increases persistently up to the high
temperature region $t \to \infty$ where it diverges according to the
square root of the temperature $\sqrt{t}$.

We have also observed a salient scaling property in the particle number
$N$ for finite temperatures, that is, the force $\Delta f(t)$ is of the order of $N$ for the bosonic case while
it is of the order of $N^2$ for the fermionic case.  Furthermore, the minimal force $\Delta
f_{\hbox{\scriptsize min}}$ and the minimal temperature
$t_{\hbox{\scriptsize min}}$ follow the same scaling law, {\it i.e.},
they are both of the order of $N$ for bosons and $N^2$ for fermions.  For  $t \to \infty$, 
the difference in the statistics disappears and 
the force 
becomes proportional to $N$ for both of the two cases.
These properties are found numerically and have also been successfully
reproduced by analytical approximations developed in this paper.   

This
curious scaling dependence pertinent to the particle statistics may
be understood intuitively by looking at the limit $t \to 0$.  Namely, in the bosonic case the force
$\Delta f(0)$ consists of the difference 
in the forces in the ground levels of the two half wells $\Delta f_1$ 
multiplied by the number $N$ of the particles all of which reside in the same levels.    In contrast, in the fermionic case the force $\Delta f(0)$ consists of the differences in the forces in the levels $\Delta f_n = e_n^- - e_n^+ = n^2 - \left(n
-1/2\right)^2 \approx n$ (see (\ref{eq:unitenergy}) and (\ref{eq:Qnfp})) 
which increase linearly up to the Fermi level $n = N$, thus yielding $\sum_{n=1}^N n \propto N^2$ in total.   For finite temperatures, this observation will be valid until the statistical property of fermions becomes obscured for higher $t$ where both of the forces tend to be proportional to $N$.
For the quantum well in one dimension, this scaling dependence pertinent to the particle statistics may also be
observed 
in other physical phenomena, not just in the quantum
force we have studied. The shift of the partition mentioned in the
previous section can be one such example.

The difference in the scaling does not necessarily imply that the quantum
force is more easily measured for the fermionic case than the bosonic
case. In fact, we have seen in the (imaginative) shift in the partition
or the redistribution of particles caused by the force that the effect of
the force in these quantities is rather marked in the bosonic case while
it is almost invisible for the fermionic case.

Compared to the bosonic case, the fermionic case admits an additional
property: for low temperatures the force exhibits a subtle step-like
pattern, a small dent, just after it starts to decrease from the constant
plateau in the extremely low temperature regime. This property has also
been reproduced by analytical approximations, at least
semiquantitatively.

{}From the expectation that all the effects that arise from quantum
boundary conditions should vanish at high temperatures where the
classical picture would be available, the observed divergence of the
force seems quite unusual. However, this may be understood by the fact
that, contrary to most quantum systems in higher dimensions, one
dimensional quantum wells possess energy spectra with increasing level
spacing for higher energy levels (which is actually valid not only for
wells with Dirichlet and/or Neumann boundary conditions but also for all
other wells as well \cite{FTC}). In other words, the dimensionality of
quantum wells can be examined by their high-temperature behaviour, too. In
actual realizations of the well, there is of course a maximal height of
the potential, which will modify the high-temperature behaviour and
eventually assimilate it to the classical one in the limit $t \to
\infty$.

It is also informative to consider what happens if the infinite well is replaced by a harmonic oscillator potential
in one dimension.  In
the harmonic case, the levels follow each other with a constant spacing.
This constant is the same on the two sides of the partition, but since the ground state energy
is different the whole spectrum on one side is shifted by a constant
with respect to the spectrum on the other.  
It turns out that the force
difference for the $n$th levels on the two sides proves to
be decreasing 
$\Delta f_n  \approx (\pi n)^{-\frac{1}{2}}$ for higher $n$ as opposed to the linearly increasing 
behaviour $\Delta f_n \approx n$ in the infinite well case. 
There are two sources for the different $n$-dependence 
in the harmonic case.  One is that the energy level difference between the two half wells is constant and does not 
give larger contributions for higher $n$.  The other is that because of the infinite stretch of the harmonic potential the
higher energy states extend more in space towards infinity and are,
correspondingly, less sensitive to a shift of the partition at the
origin.  As a result, one finds that the total net force {\it decreases} for high temperatures, with a $t^{-\frac{1}{2}}$
asymptotics.  In parallel, the $N$-dependence also changes in the harmonic
case and, for example at low temperatures, it is of the order of $N^{\frac{1}{2}}$ for
fermions (since $\sum_{n=1}^N n^{-\frac{1}{2}} \propto N^{\frac{1}{2}}$), while it remains $N$ for bosons.  Therefore, the level spacing
and the steepness of the confining potential both influence the various
aspects of the net force considerably.   The detail of the analysis for the harmonic
potential case will be reported elsewhere.

The calculation presented here could be performed for other, more general
boundary conditions as well. For most boundary conditions, however, the
energy levels will be determined by transcendental equations and,
accordingly, additional technical difficulties, especially analytical
ones, will arise. In this respect, the present combination by the
Dirichlet and Neumann conditions is certainly special in that it admits a
simple and yet distinct set of energy levels for the two half wells.
Nevertheless, one may expect that, in this example where the two
`extreme' boundary conditions are used, most of the generic features that
could arise from nontrivial boundary conditions have appeared. 

Our setting is also idealized from the point of view of possible direct
experimental verifications. For instance, inspired by
the second example mentioned
in section 6, one may consider the measurement of the instantaneous
current flow between the two half wells which is expected to occur when
they are connected by a wire. Admittedly, at the present status of
nanotechnology the boundary effects such as this may still be hard to be
examined. However, the main message of this work is that the difference
in the boundary condition creates mechanical and thermodynamical
inequilibrium, and it is a realistic assumption that future technological
developments will enable us to observe it in one form or another. We hope
that the present study furnishes an intimate picture of the effect of
boundary conditions in quantum mechanics, paving the way toward fuller
understandings of quantum singularities in general.

\ack

The research was supported in part by the Czech Ministry of Education,
Youth and Sports within the project LC06002.
Also, this work is supported by the Grant-in-Aid for Scientific Research,
No.~13135206 and No.~16540354, of the Japanese Ministry of Education,
Science, Sports and Culture.

\appendix

\section{Outline of the numerical calculation of the net force}
\label{appa}

The purely numerical computation of the net force $\Delta f$ for a given
particle number $N$ and temperature $t$ used in this paper is performed
in the following scheme. First, in each half well, we solve the total
number condition (\ref{eq:constr}) for $\alpha$, then substitute the
obtained $\alpha$ into the corresponding expression (\ref{eq:eqnQfsum})
for the force $f$, and at last we take the difference of the two forces
$f^-$ and $f^+$. We have to control errors coming from three sources. The
first is that we use finite truncations of the involved infinite sums,
the second is that we obtain $\alpha$ via a numerical solution of the
number condition, and the third is the floating-point errors of the
calculations.

For a prescribed precision of the force $f$, we estimate the required
preciseness in $\alpha$ by taking the partial derivative of the sum
(\ref{eq:eqnQfsum}) with respect to $\alpha$ at a fixed temperature
parameter $b$. Similarly, the derivative of the sum (\ref{eq:constr})
with respect to $\alpha$ is used to find the corresponding preciseness at
which (\ref{eq:constr}) has to be fulfilled when seeking for the solution
$\alpha$.

When setting this required precision, we also incorporate the error of
the truncation of the infinite sums we face at, {\it i.e.,} of
(\ref{eq:constr}), (\ref{eq:eqnQfsum}), and their mentioned derivatives. To
this end, simple upper estimate formulas for the dropped infinite terms
are derived. For low temperatures, we can estimate from above by some
appropriate geometric series
$\; \sum_{n = n_{\hbox{\scriptsize trunc}}}^\infty q^n \,\;$
or its generalization
$\; \sum_{n = n_{\hbox{\scriptsize trunc}}}^\infty
( c_2 n^2 + c_1 n + c_0 ) q^n \, ,\:$
which is summable in closed form. In parallel, for high temperatures, it
is better to replace any dropped discrete sum with a corresponding
continuous integral (cf.\  (\ref{eq:eqnQnbc})--(\ref{eq:eqnQnbd})). Then we
change the integrand to a simpler upper bound of it, to reach an integral
like $\; \int_{y_{\hbox{\scriptsize trunc}}}^\infty \ee{-y^2} \qd y \, .\:$
Finally, we employ the standard approximate closed expressions to
evaluate this simpler definite integral (the first few terms of its
large-$y_{\hbox{\scriptsize trunc}}$ expansion).

The floating-point precision must also be chosen appropriately. The
number of floating-point digits must be such high that a further increase
in the number of digits influences the force difference result well
within the uncertainty we prescribed for it. Unsurprisingly, we find that
higher temperature $t$ and larger particle number $N$ require better
floating-point precision. One should use such a mathematical computer
software that allows to have 20 digits or more. Our calculations have
been performed using Maple (\copyright\ Maplesoft, Waterloo Maple Inc.).

\section{Calculations for the medium temperature regime}
\label{appb}

For both the bosonic and the fermionic case, we will make use of an
integral approximation of the infinite sums. Namely, the trapezoid
approximation of integrals, applied for a function $g(y)$ behaving 
`peacefully' in $ [ y_1, \infty ) $ with $\lim_{y \to \infty} g(y) =
0 $, yields
  \begin{equation}\label{eq:eqnQnbc}
  \sum_{n=1}^{\infty} g(y_n) \approx \frac{g(y_1)}{2} + 
  \frac{1}{\Delta y} \int_{y_1}^{\infty} g(y) \, \qd y,
  \end{equation}
for constant intervals $ \Delta y = y_{n+1} - y_n = \hbox{const}$. This
approximation improves for smaller $\Delta y$. Remarkably, the
trapezoid approach provides an approximation one order better (in
$\Delta y$) than the simple rectangular one.

Now, in all our applications used in the text, we have
  \begin{equation}\label{eq:eqnnda}
  y_n = \sqrt{b e_n} = \sqrt{b} \, (n - \tau) \, , \hskip 7ex
  \Delta y = \sqrt{b} 
  \end{equation}
for a constant $\tau$ (defined as $\tau =  \frac{1}{2}$ for $W^+$ and
$\tau = 0$ for $W^-$ in (\ref{eq:deftau})). If $g$ is `peaceful' even in
$\, [ 0, \infty ) \, ,$ which will always be the case here, then we have
further
  \begin{eqnarray}
  \sum_{n=1}^{\infty} g(y_n) & \approx &
  \frac{g(y_1)}{2} - \frac{y_1}{\Delta y} \, g(0) + \frac{1}{\Delta y}
  \int_0^{\infty} g(y) \, \qd y \nonumber \\
  & \approx & \Big( \! \underbrace{ \frac{1}{2} - \frac{y_1}{\Delta y}
  }_{{\scriptstyle \tau - 1/2 } \atop \hbox{{\scriptsize{in our case}}}}
  \! \Big) \, g(0) + \frac{1}{\Delta y} \int_0^{\infty} f(y) \, \qd y \, .
  \label{eq:eqnQnbd}
  \end{eqnarray}
In the main body of the present paper, wherever a sum is approximated by
an integral, we have used (\ref{eq:eqnQnbd}).

Let us consider the bosonic case where we have $t \sim N$ in the medium
temperature region. There, we can show that the relative error of the
approximation (\ref{eq:eqnQnay}) tends to zero in the large-$N$ limit.
Indeed,
  \begin{eqnarray}
  & & \frac{1}{N} \sum_{n=1}^{\infty} \Big[ \underbrace{
  \frac{1}{\ee{\alpha + y_n^2} - 1} - \frac{1}{\alpha + y_n^2} }_{
  -\frac{1}{2} + \frac{\alpha + b e_n}{12} + \ordo{(\alpha + b e_n)^3} }
  \Big] \nonumber \\
  & & \approx \frac{1}{N} \left\{ \frac{1 - 2\tau}{4} \left( 1 -
  \frac{\alpha}{6} \right) + \frac{1}{ \sqrt{b} } \int_0^\infty \left(
  \frac{1}{\ee{\alpha + y^2} - 1} - \frac{1}{\alpha + y^2} \right) \qd y
  \right\}  \nonumber \\
  & & \approx \frac{1}{N} \Bigg\{ \ordo{N^0} + \ordo{N^{1/2}}
  \label{eq:eqnQnbb} \\
  & & \times \Bigg[ \underbrace{ \int_0^\infty \!
  \left( \frac{1}{\ee{y^2} - 1\vphantom{\big|}} - \frac{1}{y^2} \right)
  \qd y }_{ -1.2942 } \,t + \underbrace{ \int_0^\infty \! \left(
  \frac{1}{y^4} - \frac{\ee{y^2}} {\left( \ee{y^2} - 1 \right)^2 }
  \right) \qd y }_{ 0.1842 } \,\, \alpha \Bigg] \Bigg\} , \nonumber
  \end{eqnarray}
which is altogether an $\ordo{N^{-1/2}}$ quantity, vanishing in the
limit $N \to \infty$.

Next, we derive (\ref{eq:eqnQnbo}) and its order of relative error, under
the conditions $ b \sim 1/N \ll 1 $ and $\, |\alpha^+|, \, |\alpha^-|
\sim 1/N \ll 1 \, . $ For the force on one side, we can write
  \begin{equation}\label{eq:eqnQnbg}
  f = \sum_{n=1}^{\infty} \frac{e_n}{ \ee{ \alpha + b e_n } - 1 }  \, ,
  \hskip 3.ex  bf + N \alpha = 
  \sum_{n=1}^{\infty} \frac{\alpha + b e_n}{ \ee{\alpha + b e_n} - 1 } 
  = \sum_{n=1}^{\infty} \frac{z_n}{ \ee{z_n} - 1 }
  \end{equation}
with $ z_n := \alpha + b e_n \, . $ 
Thus, for the force difference, we have
  \begin{eqnarray}
  \Delta ( bf + N \alpha ) = b \, \Delta f + N \Delta \alpha & = &
  \sum_{n=1}^{\infty} \, \Delta \left( \frac{z_n}{ \ee{z_n} - 1 } \right)
  \nonumber \\
  & \approx & \sum_{n=1}^{\infty} \left( \frac{\qd}{\qd z} \, \frac{z}{
  \ee{z} - 1 } \right)_{ z = z_n } \cdot \Delta z_n \, .
  \label{eq:eqnQnbgg}
  \end{eqnarray}
Note that in the sum (\ref{eq:eqnQnbgg}), the factor
  \begin{equation}\label{eq:eqnQnbh}
  \Delta z_n = z_n^- - z_n^+ = \Delta \alpha + b \Delta (n - \tau)^2 =
  \Delta ( \alpha + b \tau^2 ) - 2 b \, \Delta \tau \, n
  \end{equation}
is small $ \Delta z_n = \ordo { N^{-1/2}}$ up to $\, n = \ordo{t^{1/2}}
\, , $ while the higher $n$ terms in the sum are irrelevant as being
exponentially suppressed.

In (\ref{eq:eqnQnbgg}), the term $\, \left( \frac{\qd}{\qd z} \, \frac{z}{
\ee{z} - 1 } \right)_{ z = z_n } \,$ can either be used for the quantity
in $W^+$ or $W^-$, or even for the average of the two with increased
preciseness. Fortunately, the two will prove to differ only in a
subleading order, and we do not need to specify the choice. We proceed by
rewriting $\Delta z_n$ as
  \begin{equation}\label{eq:eqnQnbj}
  \Delta z_n = [ \Delta \alpha + b ( \Delta  \tau^2 - 2 \tau 
  \Delta \tau ) ] - 2 \Delta \tau \sqrt{b} \,\, y_n
  \end{equation}
(a form that is valid on both sides $W^+$ and $W^-$), with which we
evaluate the sum (\ref{eq:eqnQnbgg}) as
  \begin{eqnarray}
  & & [ \underbrace{ \Delta \alpha + b ( \Delta  \tau^2 -
  2 \tau  \Delta \tau ) }_{ \hbox{\scriptsize{ both terms} \,}
  \ordo{N^{-1}} } ] \, \sum_{n=1}^{\infty} \left( \frac{\qd}{\qd z} \,
  \frac{z}{ \ee{z} - 1 } \right)_{ z = z_n } 
  \nonumber \\ & &
  - 2 \Delta \tau \sqrt{b} \, \sum_{n=1}^{\infty} \left( \frac{\qd}
  {\qd z} \, \frac{z}{ \ee{z} - 1 } \right)_{ z = z_n } \cdot y_n 
  \nonumber \\ & &
  \approx \ordo{N^{-1}} \bigg[ \ordo{b^0} + \frac{1}{\sqrt{b}}
  \underbrace{ \int_{y_1}^{\infty} \left( \frac{\qd}{\qd z} \, \frac{z}{
  \ee{z} - 1 } \right)_{ z = \alpha + y^2 } \qd y }_{ \ordo{b^0} } \bigg]
  \label{eq:eqnQnbk} \\ & &
  - 2 \Delta \tau \sqrt{b} \bigg[ \ordo{b^0} + \frac{1}{\sqrt{b}}
  \int_{y_1}^{\infty} \left( \frac{\qd}{\qd z} \, \frac{z}{ \ee{z} - 1 }
  \right)_{ z = \alpha + y^2 } \cdot \underbrace{y \, \qd y }_{
  \frac{1}{2} \qd z } \bigg] \, . \nonumber
  \end{eqnarray}
Among the obtained $ 2 + 2 = 4 $ terms, the first three provide only
an $\ordo{ N^{-1/2}}$ correction to the fourth term, which we
calculate as
  \begin{eqnarray}
  - \Delta \tau \int_{z_1}^{\infty} \left( \frac{\qd}{\qd z} \, \frac{z}{
  \ee{z} - 1 } \right) \qd z & = & \Delta \tau \frac{z_1}{ \ee{z_1} - 1 }
  \nonumber \\ &
  \label{eq:eqnQnbm}
  = & \Delta \tau \, [ 1 + {\cal O} \, ( \underbrace{ \alpha +
  y_1^2 }_{ \ordo{N^{-1}} } ) ] = - \frac{1}{2} + \ordo{N^{-1}} .
  \end{eqnarray}
Hence, we conclude that
\begin{equation}
  b \Delta f + N \Delta \alpha \approx - \frac{1}{2} +
  \ordo{N^{-1/2}} ,
\label{eq:noa}
\end{equation}
or
\begin{equation}
  \frac{\Delta f}{N} \approx - \frac{1}{2} \frac{t}{N} - \Delta (t \alpha) +
  \ordo{N^{-1/2}} ,
\label{eq:eqnQnbn}
\end{equation}
which is (\ref{eq:eqnQnbo}).

\end{document}